\documentclass[12pt]{article}
\usepackage[left=38mm,top=25mm,right=25mm,bottom=25mm]{geometry}
\usepackage{multicol,amssymb}
\usepackage[usenames]{color} 
\definecolor{purple}{rgb}{0.7,0,0.7}
\definecolor{LightGreen}{rgb}{0.565,0.932,0.565}
\definecolor{LightBlue}{rgb}{0.68,0.848,0.9}
\definecolor{Yellow}{rgb}{1,1,0}
\definecolor{DeepPink}{rgb}{1,0.08,0.576}

\usepackage[raggedright,large,tight,FIGTOPCAP]{subfigure} 

\usepackage{amsmath}
\usepackage{amsfonts}
\usepackage{amssymb}
\usepackage{float}
\newfloat{algorithm}{htbp}{loa}
\floatname{algorithm}{Algorithm}

\usepackage{natbib}
\newcommand{\yobs}{\mathbf{y}_{\mbox{obs}}}
\newcommand{\Sobs}{\mathbf{s}_{\mbox{obs}}}
\newcommand{\piABC}{\pi_{\mbox{\tiny{{ABC}}}}}
\newcommand{\PrABC}{\Pr_{\mbox{\tiny{{ABC}}}}}
\newcommand{\EABC}{\mbox{E}_{\mbox{\tiny{ABC}}}}
\newcommand{\VABC}{\mbox{Var}_{\mbox{\tiny{ABC}}}}

\newcommand{\x}{\mathbf{x}}
\newcommand{\s}{\mathbf{s}}
\newcommand{\Sb}{\mathbf{S}}
\newcommand{\y}{\mathbf{y}}
\newcommand{\Y}{\mathbf{Y}}

\newcommand{\ysim}{\mathbf{y}_{\mbox{sim}}}

\newtheorem{theorem}{Theorem}

\newtheorem{lem}{Lemma}

\title{Constructing Summary Statistics for Approximate Bayesian Computation: Semi-automatic ABC}

\author{ {P}aul {F}earnhead ~~%\qquad% 
and~~
{D}ennis {P}rangle
\vspace{0.2cm} 
\\
\textit{Department of Mathematics and Statistics, Lancaster
  University, UK}
}
\date{}

\usepackage{amsmath}
\usepackage{amsfonts}
\usepackage{graphicx,psfrag}

\begin{document}

\begin{centering}
\maketitle
\end{centering}

\begin{center}
{\bf Abstract}  
\end{center}

Many modern statistical applications involve inference for complex stochastic models, where it is easy to simulate from the models, but impossible to calculate likelihoods. Approximate Bayesian computation (ABC) is a method of inference for such models. It replaces calculation of the likelihood by a step which involves simulating artificial data for different parameter values, and comparing summary statistics of the simulated data to summary statistics of the observed data. Here we show how to construct appropriate summary statistics for ABC in a semi-automatic manner. We aim for summary statistics which will enable inference about certain parameters of interest to be as accurate as possible. Theoretical results show that optimal summary statistics are the posterior means of the parameters. While these cannot be calculated analytically, we  use an extra stage of simulation to estimate how the posterior means vary as a function of the data; and then use these estimates of our summary statistics within ABC.   Empirical results show that our approach is a robust method for choosing summary statistics, that can result in substantially more accurate ABC analyses than the ad-hoc choices of summary statistics proposed in the literature. We also demonstrate advantages over two alternative methods of simulation-based inference.


\section{Introduction}\label{S:Intro}

\subsection{Background}

Many modern statistical applications involve inference for stochastic models given partial observations. Often it is easy to simulate from the models but calculating the likelihood of the data, even using computationally-intensive methods, is impracticable. In these cases a natural approach to inference is to use simulations from the model for different parameter values, and to compare the simulated data sets with the observed data. Loosely, the idea is to estimate the likelihood of a given parameter value from the proportion of data sets, simulated using that parameter value, that are `similar to' the observed data. This idea dates back at least as far as \cite{Diggle/Gratton:1984}.

Note that if we replace `similar to' with `the same as' \cite[see e.g.][]{Tavare/Balding/Griffiths/Donnelly:1997}, then this approach would give an unbiased estimate of the likelihood; and asymptotically as we increase the amount of simulation we get a consistent estimate. However in most applications the probability of an exact match of the simulated data with the observed data is negligible, or zero, so we cannot consider such exact matches. The focus of this paper is how to define `similar to' for these cases.  

In this paper we focus on a particular approach: approximate Bayesian computation (ABC). This approach combines an estimate of the likelihood with a prior to produce an approximate posterior, which we will refer to as the ABC posterior. The use of ABC initially became popular within population genetics, where simulation from a range of population genetic models is possible using the coalescent \cite[]{Kingman:1982}, but where calculating likelihoods is impracticable for realistic sized data sets. The first use of ABC was by \cite{Pritchard:1999} who looked at inference about human demographic history. Further applications include inference for recombination rates \cite[]{Padhukasahasram:2006}, evolution of pathogens \cite[]{Wilson/Fearnhead:2009} and evolution of protein networks \cite[]{Ratmann:2009}. 
Its increasing importance can be seen by current range of application of ABC, which has recently been applied within epidemiology \cite[]{McKinley:2009,Tanaka:2006}, inference for extremes \cite[]{Bortot:2007}, dynamical systems \cite[]{Toni:2008} and Gibbs random fields \cite[]{Grelaud:2009} amongst many others. Part of the appeal of ABC is its flexibility, it can easily be applied to any model for which forward simulation is possible. For example \cite{Wegmann:2009} state that ABC `should allow evolutionary geneticists to reasonably estimate the parameters they are really interested in, rather than require them to shift their interest to problems for which full-likelihood solutions are available'. Recently there has been software developed to help implement ABC within population genetics \cite[]{Cornuet:2008,Lopes:2009}, and systems biology \cite[]{Liepe:2010}

%For concreteness we will focus on one specific implementation of ABC, which uses importance sampling (or as a special case, rejection sampling). There are other approaches to implementing ABC, for example using MCMC \cite[]{Marjoram:2003,Bortot:2007} or sequential importance sampling \cite[]{Beaumont:2009,Sisson:2007}. The issue we focus on, that of how to define when simulated data is sufficiently similar to the observed data, is common to all methods -- and we believe the results we obtain, and the resulting guidance for implementing ABC, are applicable to all ABC methods. 

\subsection{ABC Algorithms and Approximations} \label{Sec:1.2}

Consider analysing $n$-dimensional data  $\yobs$. We have a model for the data, which depends on an unknown $p$-dimensional parameter, $\theta$. Denote the probability density of the data given a specific parameter value by $\pi(\y|\theta)$, and denote our prior by $\pi(\theta)$. We assume it is simple to simulate $\Y$ from $\pi(\y|\theta)$ for any $\theta$, but that we do not have an analytic form for $\pi(\y|\theta)$. 

We define the ABC posterior in terms of (i) a function $S(\cdot)$ which maps the $n$-dimensional data onto a $d$-dimensional summary statistic; (ii) a density kernel $K(\x)$ for a $d$-dimensional vector $\x$, which integrates to 1; and (iii) a bandwidth $h>0$. Let $\Sobs=S(\yobs)$.  If we now define an approximation to the likelihood as
\[
p(\theta|\Sobs)=\int \pi(\y|\theta) K(\{S(\y)-\Sobs\}/h)\mbox{d}\y,
\]
then the ABC posterior can be defined as
\begin{equation} \label{eq:1a}
\piABC(\theta|\Sobs)\propto{\pi(\theta)p(\theta|\Sobs)}.
\end{equation}
The idea of ABC is that the ABC posterior will approximate, in some way, the true posterior for $\theta$ and can be used for inference about $\theta$. The form of approximation will depend on the choice of $S(\cdot)$, $K(\cdot)$ and $h$. For example, if $S(\cdot)$ is the identity function then we can view $p(\theta|\Sobs)$ as a kernel density approximation to the likelihood. If also $h\rightarrow 0$, then the ABC posterior will converge to the true posterior. For other choices of $S(\cdot)$, the kernel is measuring the closeness of $\y$ to $\yobs$ just via the closeness of $S(\y)$ to $\Sobs$.
The reason for considering ABC is that we can construct Monte Carlo algorithms which approximate the ABC posterior and only require the
ability to be able to simulate from $\pi(\y|\theta)$. 

For simplicity of the future exposition, we will assume that $\max K(\x)=1$. This assumption imposes no restriction, as if $K_0(\x)$ is a density kernel, then so is $K(\x)=h_0^{-d} K_0(\x/h_0)$ for any $h_0>0$. Thus we can choose $h_0$ so that $\max K(\x)=1$. Note that a bandwidth $\lambda$ for kernel $K_0(\x)$ is equivalent to a bandwidth $h=\lambda h_0$ for $K(\x)$, so the value of $h_0$ just redefines the units of the bandwidth $h$.

One algorithm for approximating the ABC posterior (\ref{eq:1a}), based upon importance sampling, is given by Algorithm \ref{Alg:1}. \cite[For alternative importance sampling approaches, based on sequential Monte Carlo, see][]{Beaumont:2009,Sisson:2007,Sisson:2009}. A standard importance sampler for approximating the ABC posterior would repeatedly simulate a paramater value, $\theta'$, from a proposal $g(\theta)$ and then assign it an importance sampling weight
\begin{equation} \label{eq:ISex}
\frac{\pi(\theta')}{g(\theta')} \int \pi(\y|\theta') K(\{S(\y)-\Sobs\}/h)\mbox{d}\y 
\end{equation}
This is not possible here, as $\pi(\y|\theta)$ is intractable, so Algorithm \ref{Alg:1} introduces an extra Monte Carlo step. This step first simulates $\ysim$ from $\pi(\y|\theta')$ and then accepts this value with probability $K(\{S(\ysim)-\Sobs\}/h)$. Accepted values are assigned weights $\pi(\theta')/g(\theta')$. The key thing is that the expected value of such weights is just (\ref{eq:ISex}), which is all that is required for this to be a valid importance sampling algorithm targeting (\ref{eq:1a}).

The output of Algorithm \ref{Alg:1} is a weighted sample of $\theta$ values, which approximates the ABC posterior of $\theta$. Many of the weights will be 0, and in practice we would remove the corresponding $\theta$ values from the sample. A specific case of Algorithm \ref{Alg:1} occurs when $g(\theta)=\pi(\theta)$, and then we have a rejection sampling algorithm. The most common implementation of ABC has a deterministic accept-reject decision in step 3, and this corresponds to $K(\cdot)$ being the density of a uniform random variable -- the support of the uniform random variable defining the values of $\bf{s}-\Sobs$ which are accepted. 

\begin{algorithm}[ht]
\begin{footnotesize}
\rule{\textwidth}{1mm}
\underline{Importance Sampling ABC}

\begin{tabular}[h]{ll}
{\bf Input:} & A set of data $\yobs$, and a function $S(\cdot)$. \\
& A density kernel $K(\cdot)$, with $\max K(\x) =1$ and a bandwidth $h>0$. \\
& A proposal density $g(\theta)$; with  $g(\theta)>0$ if $\pi(\theta)>0$.\\
& An integer $N>0$.\\
\\
{\bf Initialise:} & Define $\Sobs=S(\yobs)$.
\\
\\ 
\end{tabular}

{\bf Iterate:} For $i=1,\ldots,N$
\begin{enumerate}
\item Simulate $\theta_i$ from $g(\theta)$.
\item Simulate $\ysim$ from $\pi(\y|\theta_i)$, and calculate $\s=S(\ysim)$.
\item With probability $K( (\s-\Sobs)/h$ set $w_i=\pi(\theta_i)/g(\theta_i)$, otherwise set $w_i=0$.
\end{enumerate}

{\bf Output:} A set of parameter values $\{\theta_i\}_{i=1}^N$ and corresponding weights $\{ w_i \}_{i=1}^N$.

\rule{\textwidth}{1mm}
\caption{\footnotesize Importance (and rejection) sampling implementation of ABC. \label{Alg:1}}
\end{footnotesize}
\end{algorithm}

An alternative Monte Carlo procedure for implementing ABC is based on MCMC \cite[]{Marjoram:2003,Bortot:2007}, and given in Algorithm \ref{Alg:3}. Both this and Algorithm \ref{Alg:1} target the same ABC posterior \cite[for a proof of the validity of this algorithm see][]{Sisson:2010}. 

\begin{algorithm}[ht]
\begin{footnotesize}
\rule{\textwidth}{1mm}
\underline{MCMC ABC}

\begin{tabular}[h]{ll}
{\bf Input:}& A set of data $\yobs$, and a function $S(\cdot)$. \\
& A density kernel $K(\cdot)$, with $\max K(\x) =1$ and a bandwidth $h>0$. \\
& A transition kernel $g(\cdot|\cdot)$.\\
& An integer $N>0$.\\

\\
{\bf Initialise:} & Define $\Sobs=S(\yobs)$, and choose or simulate $\theta_0$ and $\s_0$.
\\
\\ 
\end{tabular}

{\bf Iterate:} For $i=1,\ldots,N$
\begin{enumerate}
\item Simulate $\theta$ from $g(\theta|\theta_{i-1})$.
\item Simulate $\ysim$ from $\pi(\y|\theta)$, and calculate $\s=S(\ysim)$.
\item With probability 
\[
\min \left\{1,\frac{K( (\s-\Sobs)/h)}{K((\s_{i-1}-\Sobs)/h)}\frac{\pi(\theta)g(\theta_{i-1}|\theta)}{\pi(\theta_{i-1})g(\theta|\theta_{i-1})}\right\} 
\]
accept $\theta$ and set $\theta_i=\theta$ and $\s_i=\s$; otherwise set $\theta_i=\theta_{i-1}$ and $\s_i=\s_{i-1}$.
\end{enumerate}

{\bf Output:} A set of parameter values $\{\theta_i\}_{i=1}^N$.

\rule{\textwidth}{1mm}
\caption{\footnotesize MCMC sampling implementation of ABC. \label{Alg:3}}
\end{footnotesize}
\end{algorithm}

Good implementation of ABC requires a trade-off between the approximation error between the ABC posterior and the true posterior, and the Monte Carlo approximation error of the ABC posterior. The last of these will also be affected by the algorithm used to implement ABC, and the specific implementation of that algorithm. We will consider both of these approximations in turn.

To understand the former approximation, consider the following alternative expression of the ABC posterior as a continuous mixture of posteriors. Consider the random variable $\Sb=S(\Y)$, which is just the summary statistic of the data. Denote the posterior for $\theta$ given $\Sb=\s$ by $\pi(\theta|\s)$, and the marginal density for $\Sb$ by $\pi(\s):=\int \pi(\s|\theta)\pi(\theta)\mbox{d}\theta$. Then, 
\begin{equation} \label{eq:3}
\piABC(\theta|\Sobs)= \int \beta(\s) \pi(\theta|\s) \mbox{d}\s, \mbox{ where } \beta(\s)=\frac{\pi(\s) K(\{\s-\Sobs\}/h)}{\int \pi(\s)K(\{\s-\Sobs\}/h) \mbox{d}\s}.
\end{equation}
The mixing weight for a given value of $\Sb$ is just the conditional density for such a value of $\Sb$ given acceptance in step 3 of Algorithm \ref{Alg:1}. If $h$ is small then  $\piABC(\theta|\Sobs)\approx \pi(\theta|\Sobs)$. Thus we can split the approximation of the posterior by ABC into the approximation of $\pi(\theta|\yobs)$ by $\pi(\theta|\Sobs)$, and the further approximation of $\pi(\theta|\Sobs)$ by $\piABC(\theta|\Sobs)$. The former is controlled by the choice of $S(\cdot)$, the latter by $K(\cdot)$ and $h$. 

Now consider the Monte Carlo approximation within  importance sampling ABC. %The approximation of $\piABC(\theta|\Sobs)$ by the weighted sample $\{\theta_i,w_i\}_{i=1}^N$ can be defined in terms of estimates of the ABC posterior means of functions of $\theta$ as follows. 
Consider a scalar function $h(\theta)$, and define the ABC posterior mean of $h(\theta)$ as
\[
\EABC(h(\theta)|\Sobs) = \int h(\theta)\piABC(\theta|\Sobs) \mbox{d}\theta.
\]
Assuming this exists then, by the law of large numbers, as $N\rightarrow\infty$
\begin{equation} \label{eq:1}
%\hEABC(h(\theta)|\yobs):=
\frac{\sum_{i=1}^N w_ih(\theta_i)}{\sum_{i=1}^N w_i} \rightarrow \EABC(h(\theta)|\Sobs),
\end{equation}
where convergence is in probability.
%%%
%%% Variance calculations complicated by the ratio. 
%%%
For rejection sampling, for large $N$ the variance of this estimator is just
\begin{equation} \label{eq:2}
\frac{\VABC (h(\theta)|\Sobs)}{N_{\mbox{acc}}},
\end{equation}
where the numerator is the ABC posterior variance of $h(\theta)$, and the denominator is
$N_{\mbox{acc}}=N \int p(\theta|\Sobs)\pi(\theta) \mbox{d}\theta$, the expected number of acceptances. 

Similar calculations for both importance sampling in general and for the MCMC version of ABC are given in Appendix A. 
In all cases the Monte Carlo error depends on $\int p(\theta|\Sobs)\pi(\theta) \mbox{d}\theta$, the average acceptance probability of the rejection sampling algorithm. %This will be used as a proxy for the Monte Carlo error in Section \ref{S:KER}.
The following Lemma characterises this acceptance probability for small $h$ in this case of continuous summary statistics (similar results can be shown for discrete summary statistics). 
\begin{lem} \label{l:1}
Assume that either (i) the marginal distribution for the summary statistics, $\pi(\s)$ is continuous at $\s=\Sobs$ and that the
kernel $K(\cdot)$ has finite support; or (ii) $\pi(\s)$ is continuously differentiable, $| \partial \pi(\s)/\partial s_i|$ is bounded above for all $i$, and $\int |\x_i|K(\x)\mbox{d}\x$ is bounded for all $i$. Then, in the limit as $h\rightarrow 0$  
\begin{equation} \label{eq:4}
\int p(\theta|\Sobs)\pi(\theta) \mbox{d}\theta = \pi(\Sobs) h^{d} + o\left( h^{d} \right),
\end{equation}
where $d$ is the dimension of $\Sobs$. 
\end{lem}
{\bf{Proof:}} See Appendix B. \\

This result gives insight into how $S(\cdot)$ and $h$ affect the Monte Carlo error. To minimise Monte Carlo error, we need $h^d$ to not be too small.
Thus ideally we want $S(\cdot)$ to be a low-dimensional summary of the data that is sufficiently informative about $\theta$ so that $\pi(\theta|\Sobs)$ is close, in some sense, to $\pi(\theta|\yobs)$. The choice of $h$ affects the accuracy of $\piABC$  in approximating $\pi(\theta|\Sobs)$, but also the average acceptance probability (\ref{eq:4}), and hence the Monte Carlo error. 

\subsection{Approach and Outline}

Previous justification for ABC has, at least informally, been based around $\piABC$ approximating $\pi(\theta|\yobs)$ globally. This is possible if $\yobs$ is low-dimensional, and $S(\cdot)$ is the identity function, or if we have low-dimensional sufficient statistics for the data. In these cases we can control the error in the ABC approximation by choosing $h$ sufficiently small. In general applications this is not the case. Arguments have been made about trying to choose approximate sufficient statistics \cite[see e.g.][]{Joyce/Marjoram:2008}. However the definition of such statistics is not clear, and more importantly it is difficult or impossible to construct a general method for finding such statistics.

We take a different approach, and weaken the requirement for $\piABC$ to be a good approximation to $\pi(\theta|\yobs)$. We argue for $\piABC$ to be a good approximation solely in terms of the {\it accuracy of certain estimates} of the parameters.
We also would like to know how to interpret the ABC posterior and probability statements derived from it. To do this we consider a property we call {\em calibration}, which we define formally below. If $\piABC$ is calibrated, then this means that probability statements derived from it are appropriate, and in particular that we can use $\piABC$ to quantify uncertainty in estimates. We argue that using such criteria we can construct ABC posteriors which have both good inferential properties, and which can be estimated well using Monte Carlo methods such as in Algorithm \ref{Alg:1}. %Considering both accuracy and calibration has links to approaches in probabilistic forecasting \cite[]{Gneiting:2008}

In Section \ref{S:CA} we define formally what we mean be calibration and accuracy. Standard ABC is not calibrated, but a simple modification, which we call {\it noisy ABC}, is. We further show that if the bandwidth, $h$, is small then standard ABC is approximately calibrated. Theoretical results are used to produce recommendations about when standard ABC and when noisy ABC should be used. Then, for a certain definition of accuracy we show that the optimal choice of summary statistics $S(\Y)$  are the true posterior means of the parameters. Whilst these are unknown, simulation can be used to estimate the summary statistics. Our approach to doing this is described in Section \ref{S:Alg}, together with results that show the advantage of ABC over directly using the summary statistics to estimate parameters. Section \ref{S:Ex} gives examples comparing our implementation of ABC with previous methods. The paper ends with a discussion.

\section{Calibration and Accuracy of ABC} \label{S:CA}

Firstly we define what we mean by calibration, and introduce a version of ABC that is calibrated. We show how the idea of calibration can be particularly important when analysing multiple data sets. We then discuss our definition of accuracy of the ABC posterior, and how this can be used to guide the choice of summary statistic we use.

\subsection{Calibration and Noisy ABC}

Consider a subset of the parameter space $\mathcal{A}$.  For given data $\yobs$, the ABC posterior will assign a probability to the event $\theta \in \mathcal{A}$
\[
\PrABC(\theta \in \mathcal{A}|\Sobs)=\int_{\mathcal{A}} \piABC(\theta|\Sobs)\mbox{d}\theta.
\]
For a given probability, $q$, consider the event $E_q(\mathcal{A})$ that $\PrABC(\theta \in \mathcal{A}|\Sobs)=q$. Then the ABC posterior is calibrated if
\begin{equation} \label{eq:6}
\Pr(\theta \in \mathcal{A}|E_q(\mathcal{A}))=q.
\end{equation}
%In defining the probability, we allow for any randomness within the construction of $\Sobs$ (see below).
The probability is then defined in terms of the density on parameters and data given by our prior and likelihood model,  $\pi(\theta)\pi(\y|\theta)$, but ignoring any Monte Carlo randomness. 
Statement (\ref{eq:6}) states that under repeated sampling from the prior, data, and summary statistics, events assigned probability $q$ by the ABC posterior will occur with probability $q$. A consequence of calibration is that the ABC posterior will appropriately represent uncertainty in the parameters: for example we can construct appropriate credible intervals. That is calibration means we can use the ABC posterior as we would any standard posterior distribution.

Standard ABC posteriors (\ref{eq:3}) are not calibrated in general. Instead
we introduce the idea of {\it noisy ABC} which is calibrated. Noisy ABC involves defining summary statistics which are random. A noisy ABC importance sampling algorithm is obtained by changing the initialisation within Algorithm \ref{Alg:1} or Algorithm \ref{Alg:3}; details are given in Algorithm \ref{Alg:2}. The resulting ABC posterior for a given $\yobs$ is random, and the definition of the probability in (\ref{eq:6}) needs to account for this extra randomness. 

\begin{algorithm}[ht]
\begin{footnotesize}
\rule{\textwidth}{1mm}
\underline{Noisy ABC}\\
Replace the initialisation step in Algorithm \ref{Alg:1} and \ref{Alg:3} with: \\
{\bf Initialise:} Simulate $\x$ from $K(\x)$. Define $\Sobs=S(\yobs)+h\x$.\\
\rule{\textwidth}{1mm}
\caption{\footnotesize Change of Algorithm \ref{Alg:1} or \ref{Alg:3} for noisy ABC. \label{Alg:2}}
\end{footnotesize}
\end{algorithm}

\begin{theorem} \label{thm:NABC}
Algorithm \ref{Alg:2} produces an ABC posterior that is calibrated.
\end{theorem}
{\bf{Proof:}} The ABC posterior derived by Algorithm \ref{Alg:2} is $\piABC(\theta|\Sobs)$ where $\Sobs$ is related to the data by $\yobs$ by
\begin{equation} \label{eq:7}
\Sobs=S(\yobs)+h\x,
\end{equation}
and $\x$ is the realisation of a random variable with density $K(\x)$. However the definition of $\piABC(\theta|\Sobs)$ is just that of the true posterior for $\theta$ given data $\Sobs$ generated by (\ref{eq:7}). It immediately follows that this density is calibrated. \hfill $\Box$

%An algebraic proof is given in \cite{Prangle:2010}. 
A related idea is used by \cite{Wilkinson:2008} who shows that the ABC posterior is equivalent to the true posterior under an assumption of appropriate model error. In the limit as $h \rightarrow 0$ noisy ABC is equivalent to standard ABC, we discuss the links between the two in more detail in Section \ref{S:Links}. 

\subsection{Inference from Multiple Data Sources} \label{S:mult}

Consider combining data from $m$ independent sources, $\yobs^{(1)},\ldots,\yobs^{(m)}$. It is possible to use individual ABC analyses for each data set, and then combine these inferences. One sequential approach is to use the ABC posterior after analysing the $i$th data set as a prior for analysing the $(i+1)$st data set. Algorithms for such an approach are a special case of those discussed in \cite{Wilkinson:2011}. 

One consequence of calibration is that such inferences will be well behaved in the limit as $m$ gets large. To see this define the ABC approximation to the $i$th data set as $p(\theta|\Sobs^{(i)})$, and note that the above approach is targeting the follow ABC posterior
\[
 \piABC(\theta|\Sobs^{(1)},\ldots,\Sobs^{(m)}) \propto \pi(\theta)\prod_{i=1}^m p(\theta|\Sobs^{(i)}).
\]
If we use noisy ABC, where $\Sobs^{(i)}$ is random centered on $S(\yobs^{(i)})$, then by same argument as for Thereom \ref{thm:NABC} this ABC posterior will be calibrated. Furthermore, we have the following result
\begin{theorem} \label{thm:5}
Let $\theta_0$ be the true parameter value. 
Consider noisy ABC, where $\Sobs=S(\yobs)+h\x$, where $\x$ is drawn from $K(\cdot)$, then the expected noisy-ABC log-likelihood,
\[
\mbox{E}\left\{ \log[p(\theta|\Sb_{\mbox{obs}})]  
\right\} = \int \int \log[p(\theta|\Sb(\y)+h\x)]\pi(\y|\theta_0)K(\x)\mbox{d}\y\mbox{d}\x ,
\]
has its maximum at $\theta=\theta_0$.
\end{theorem}
{\bf{Proof:}} Make the change of variable $\Sobs=\Sb(\y)+h\x$, then 
\[
\mbox{E}\left\{ \log[p(\theta|\Sb_{\mbox{obs}})]  
\right\} = \frac{1}{h}\int \int \log[p(\theta|\Sobs)]\pi(\y|\theta_0)K( [\Sobs-\Sb(\y)]/h)\mbox{d}\y\mbox{d}\Sobs.
\]
Now by definition $p(\theta|\Sobs)=\int \pi(\y|\theta)K( [\Sobs-\Sb(\y)]/h)\mbox{d}\y$, thus we get
\[
 \mbox{E}\left\{ \log[p(\theta|\Sb_{\mbox{obs}})]  
\right\} = \frac{1}{h}\int \int \log[p(\theta|\Sobs]p(\theta|\Sobs) \mbox{d}\Sobs.
\]
and by Jensen's inequality this has its maximum at $\theta=\theta_0$. \hfill $\Box$

The importance of this result, is that under the standard regularity conditions \cite[]{Bernardo/Smith:1994}, the ABC posterior will converge onto a point mass on the true parameter value as $m\rightarrow \infty$. By comparison, if we use standard ABC, then we have no such guarantee. As a simple example assume $Y^{(i)}$ are independent identically distributed from a Normal distribution with mean 0 and variance $\sigma^2$, and our kernel is chosen to be Normal with variance $\tau<\sigma^2$. If we use standard ABC, the ABC posterior given $y^{(1)},\ldots,y^{(m)}$ will converge to a point mass on $\sigma^2-\tau$ as $m\rightarrow \infty$.

We look empirically at this issue in Section \ref{S:ExSKN}.

\subsection{Accuracy and Choice of Summary Statistics} \label{sec:SS}

Calibration itself is not sufficient to define a sensible ABC posterior. For example the prior distribution is always calibrated, but will not give accurate estimates of parameters. Thus we also want to maximise accuracy of estimates based on the ABC posterior. We will define accuracy in terms of a loss function for estimating the parameters. A natural choice of loss function is quadratic loss. Let $\theta_0$ be the true parameter values, and $\hat{\theta}$ an estimate. Then we will consider the class of loss functions, defined in terms of a $p\times p$ positive-definite matrix $A$,
\begin{equation} \label{eq:loss}
L(\theta_0,\hat{\theta};A)= (\theta_0-\hat{\theta})^T A (\theta_0-\hat{\theta}).
\end{equation}
We now consider implementing ABC to minimise this quadratic error loss of estimating the parameters. We consider the limit of $h\rightarrow 0$. This will give results that define the optimal choice of summary statistics, $S(\cdot)$.

For any choice of weight matrix $A$ that is of full-rank, the following theorem shows that the optimal choice of summary statistics is $S(\yobs)=\mbox{E}(\theta|\yobs)$, the true posterior mean.
\begin{theorem} \label{thm:1}
Consider a $p \times p$ positive-definite matrix $A$ of full rank. Given observation $\yobs$, let $\Sigma$ be the true posterior variance for $\theta$. Then
\begin{itemize}
\item[(i)] The minimal possible quadratic error loss, $E(L(\theta,\hat{\theta};A)|\yobs)$ occurs when $\hat{\theta}=\mbox{E}(\theta|\yobs)$ and is $\mbox{trace}(A\Sigma)$.
\item[(ii)] If $S(\yobs)=\mbox{E}(\theta|\yobs)$ then in the limit as $h\rightarrow 0$ the minimum loss, based on inference using the  ABC posterior, is achieved by $\hat{\theta}=\EABC(\theta|\Sobs)$. The resulting expected loss is $\mbox{trace}(A\Sigma)$.
\end{itemize}
\end{theorem}
{\bf{Proof:}} Part (i) is a standard result of Bayesian statistics \cite[]{Bernardo/Smith:1994}. For (ii) we just need to show that in the limit as $h\rightarrow 0$, $\EABC(\theta|\Sobs)=\mbox{E}(\theta|\yobs)$. By definition in the limit $h\rightarrow 0$ $\Sobs=S(\yobs)$ with probability 1, and $\piABC(\theta|\Sobs)=\pi(\theta|\Sobs)$. Furthermore
\begin{eqnarray*}
\EABC(\theta|\Sobs) &=& \int \theta\pi(\theta|\Sobs)\mbox{d}\theta, \\
&=& \int \int \theta \pi(\theta|\y) \pi(\y|\Sobs)\mbox{d}\y\mbox{d}\theta, 
\end{eqnarray*}
where $\pi(\y|\Sobs)$ is the conditional distribution of the data $\y$ given the summary statistic $\Sobs$. Finally by definition, all $\y$ that are consistent with $\Sobs$ satisfy $\int \theta \pi(\theta|\y) \mbox{d}\theta=\mbox{E}(\theta|\yobs)$, and hence the result follows. \hfill $\Box$

Use of square error loss leads to  ABC approximations that attempt to have the same posterior mean as the true posterior. Using alternative loss functions would mean matching other features of the posterior: for example absolute error loss would result in matching the posterior medians.
It is possible to choose other summary statistics that also achieve the minimum expected loss. However any such statistic with dimension $d>p$ will cause larger Monte Carlo error (see Lemma \ref{l:1} and the discussion below).

\subsection{Comparison of standard ABC and noisy ABC} \label{S:Links}

Standard and noisy ABC are equivalent in the limit as $h\rightarrow0$, with the ABC posteriors converging to $\mbox{E}(\theta|\Sb(\yobs))$. We can further quantify the accuracy of estimates based on standard or noisy ABC for $h\approx 0$. For noisy ABC we have the following result
\begin{theorem} \label{thm:2}
Assume condition (i) of Lemma \ref{l:1}, that $\pi(\mbox{E}(\theta|\yobs)) > 0$, and the kernel $K(\cdot)$ corresponds to a random variable with mean zero.  If $\Sb(\yobs)=\mbox{E}(\theta|\yobs)$ then for small $h$ the expected quadratic loss associated with $\hat{\theta}=\EABC(\theta|\Sobs)$ is
\[
\mbox{E}(L(\theta,\hat{\theta};A)|\yobs)=\mbox{trace}(A\Sigma)+h^2\int \x^T A \x K(\x)\mbox{d}\x+o(h^2).
\]
\end{theorem}
{\bf{Proof:}} The idea is that $\EABC(\theta|\Sobs) = \Sobs+o(h)$; and the squared error loss based on $\hat{\theta}=\Sobs$ is just $\mbox{trace}(A\Sigma)+h^2\int \x^T A \x K(\x)\mbox{d}\x$. See Appendix C.

A similar result exists for standard ABC \cite[]{Prangle:2010}, which shows that in this case
\[
\mbox{E}(L(\theta,\hat{\theta};A)|\yobs)=\mbox{trace}(A\Sigma)+O(h^4).% \pi(\Sobs)^{-2} \textbf{v}^T A \textbf{v} + O(h^5),
\]
Extensions of these results can be used to give guidance on the choice of kernel. For noisy ABC they suggest a uniform kernel on the ellipse $\x^T A \x <c$, for some $c$, for standard ABC they also suggest a uniform kernel on an ellipse, but the form of the ellipse is difficult to calculate in practice. We do not give these results in more detail, as in practice we have found the choice of kernel has relatively little effect on the accuracy of either ABC algorithm.

These two results also give an insight into the overall accuracy of a Monte Carlo ABC algorithm. For simplicity consider a rejection sampling algorithm. The Monte Carlo variance is inversely proportional to the acceptance probability. Thus using Lemma \ref{l:1} we have that the Monte Carlo variance is $O(N^{-1}h^{-d})$, where $N$ is the number of proposals. The expected quadratic loss based on estimates from the ABC importance sampler will been increased by this amount. Thus for noisy ABC we want to choose $h$ to minimise
\[
\mbox{trace}(A\Sigma)+h^2\int \x^T A \x K(\x)\mbox{d}\x+\frac{C_0}{Nh^{d}},
\]
for some constant $C_0$. This gives that we want $h=O(N^{-1/(2+d)})$, and the overall expected loss above $\mbox{trace}(A\Sigma)$ would then decay as $N^{-2/(2+d)}$. For standard ABC a similar argument give $h=O(N^{-1/(4+d)})$, and the overall expected loss would then decay as $N^{-4/(4+d)}$.

Thus we can see that the choice between using standard ABC and noisy ABC is one of a trade-off between accuracy and calibration. Noisy ABC is calibrated, but for small $h$ will give less accurate estimates. As such for the analysis of a single data set where the number of summary statistics is not too large, and hence $h$ is small, we would recommend the use of standard ABC. If we wish to combine inferences from ABC analyses of multiple data sets, then in light of the discussion in Section \ref{S:mult}, we would recommend noisy ABC. For all the examples in Section \ref{S:Ex} we found that this approach worked well in practice.

Note that possibly the best approach is to use noisy ABC, but to use Rao-Blackwellisation ideas to average out the noise that is added to the summary statistics. Such an approach would have the guarantee that the resulting expected quadratic loss for estimating any function of the parameters would be smaller than that from noisy ABC. However implementing such a Rao-Blackwellisation scheme efficiently appears non-trivial, with the only simple approach being to run noisy ABC independently on the same data set, and then to  average the estimates across each of these runs. 

\section{Semi-automatic ABC} \label{S:Alg}

%ADAPT TEXT FROM UPGRADE ABOUT IMPLEMENTING ABC. INCLUDE CHOICE OF $A$. SOME DETAILS ON RELATIVE COMPUTATION TIME. DESCRIBE THE CHOICE OF METHOD TO ESTIMATE THE POSTERIOR MEANS LATER. I.E. INITIALLY HAVE AN ALGORITHM WHICH USES SIMULATION TO FIND THEM (BUT DO NOT SPECIFY HOW). NEED TO EXPLAIN/DESCRIBE THE PILOT STUDY -- PARTICULARLY IMPORTANT IF YOU HAVE DIFFUSE PRIORS.

The above theory suggests that we wish to choose summary statistics that are equal to posterior means. Whilst we cannot use this result directly, as we cannot calculate the posterior means, we can use simulation to estimate appropriate summary statistics.

Our approach is to:
\begin{itemize} 
 \item[(i)] use a pilot run of ABC to determine a region of non-negligible posterior mass;
 \item[(ii)] simulate sets of parameter values and data; 
\item[(iii)] use the simulated sets of parameter values and data to estimate the summary statistics; and
\item[(iv)] run ABC with this choice of summary statistics.
\end{itemize}

Step (i) of this algorithm is optional. Its aim is to help define an appropriate training region of parameter space from which we should simulate parameter values. In applications where our prior distributions are relatively informative, this step should be avoided as we can simulate parameter values from the prior in step (ii). However it is important if we have uninformative priors, particularly if they are improper.

If we implement (i), we assume we have arbitrarily chosen some summary statistics to use within ABC. In our implementation below we choose our training region as a hypercube, with the range for each parameter being the range of that parameter observed within our pilot run. Then in (ii) we simulate parameter values from the prior truncated to this training region, and for each choice of parameter value we simulate an artificial data set. We repeat this $M$ times, so that we have $M$ sets of parameter values, each with a corresponding simulated data set. 

There are various approaches that we can take for (iii). In practice we found that using linear regression, with appropriate functions of the data as predictors, is both simple and  worked well. We also considered using LASSO \cite[]{Hastie:2001}, and canonical correlation analysis \cite[]{Mardia:1979} but in general neither of these performed better than linear regression (though the LASSO maybe appropriate if we wish to use a large number of explanatory variables within the linear model).

Our linear regression approach involved considering each parameter in term. First we introduce a vector-valued function $f(\cdot)$, so that $f(\y)$ is a vector of, possibly non-linear, transformations of the data. The simplest choice is $f(\y)=\y$, but in practice including other or different transformation as well may be beneficial. For example, in one application below we found $f(\y)=(\y,\y^2,\y^3,\y^4)$, that is a vector of length $4n$ that consists of the data plus all second to fourth powers of individual data points, produced a better set of summary statistics.

For the $i$th summary statistic
the simulated  values of the $i$th parameter, $\theta_i^{(1)}, \ldots, \theta_i^{(M)}$, are used as the responses; and 
the transformations of the simulated data, $f(\y^{(1)}),\ldots,f(\y^{(M)})$, are used as the explanatory variables. We then fit the model
\[
 \theta_i=\mbox{E}(\theta_i|\y)+\epsilon_i = \beta_0^{(i)}+\beta^{(i)}f(\y)+\epsilon_i,
\]
where $\epsilon_i$ is some mean-zero noise, using least squares. The fitted function $\hat{\beta}_0^{(i)}+\hat{\beta}^{(i)}f(\y)$ is then an estimate of $\mbox{E}(\theta_i|\y)$. The constant terms can be neglected in practice as ABC only uses the difference in summary statistics. Thus the $i$th summary statistic for ABC is just $\hat{\beta}^{(i)}f(\y)$.

Note that our approach of using a training region means that our model for the posterior means are based only on parameter values simulated within this region. We therefore suggest adapting the ABC run in step (iv) so that the prior is truncated to lie within this training region \cite[a similar idea is used in][]{Blum/Francois:2010}. This can be viewed as using, weakly, the information we have from the pilot ABC run within the final ABC run, and has links with composite likelihood methods \cite[]{Lindsay:1988}. More importantly it makes the overall algorithm robust to problems where $\mbox{E}(\hat{\beta}^{(i)}f(\Y)|\theta_i)$ is similar for  two dissimilar values of $\theta_i$, one inside the training region and one outside. 

In practice below we use roughly a quarter of our total CPU time on (i) and (ii) and half on (iv), with step (iii) having negligible CPU cost. Note that we call this semi-automatic ABC as the choice of summary statistics is now based on simulation, but that there are still choices by the user in terms of fitting the linear model in step (iii). This input is in terms of the choice of $f(\y)$ to use.
Note that (iii) is now a familiar statistical problem, and that standard model checks can be used to decide whether that choice of $f(\y)$ is appropriate, and if not, how it could be improved. Also note that repeating step (iii) with a different choice of $f(\y$) can be done without any further simulation of data, and thus is quick in terms of the CPU cost. Furthermore standard model comparison procedures (e.g. using BIC) can be used to choose between summary statistics obtained from linear-regressions using different explanatory variables.

A natural question is whether this approach is better than the current approach to ABC, where summary statistics are chosen arbitrarily. In our implementation we still need to choose summary statistics for step (i), and we also need to choose the set of explanatory variables for the linear model. However we believe our approach is more robust to these choices than standard ABC. Firstly the choice of summary statistics in step (i) is purely to make step (ii) more efficient, and as such the final results depend little on this choice. Secondly, when we choose the explanatory variables we are able to choose many such variables (of the order of 100s). As such we are much more likely to include amongst these some variables which are informative about the  parameters of interest than standard ABC where generally a few summary statistics are used. If many summary statistics are used in ABC, then this will require a large value of $h$, and will often be inefficient due to the accept/reject decision being based not only on the informative summary statistics, but also those which are less informative. These issues are demonstrated empirically in the examples we consider.

Our approach has similarities to that of \cite{Beaumont:2002} \cite[see also][]{Blum/Francois:2010}, where the authors use linear regression to correct the output from ABC.
The key difference is that our approach uses linear regression to construct the summary statistics, whereas \cite{Beaumont:2002} uses linear regression to reduce the error between $\piABC(\theta|\Sobs)$ and $\pi(\theta|\Sobs)$. In particular the method of \cite{Beaumont:2002} assumes that appropriate low-dimensional summary statistics have already been chosen. %As such there is the possibility of using the idea of \cite{Beaumont:2002} on the output from our ABC algorithm. This could be particularly important for models with many parameters, and hence many summary statistics. 
%A recent paper, \cite{Blum/Francois:2010}, extends the idea of \cite{Beaumont:2002} by using non-linear regression techniques, and also using a two-stage procedure similar to the one we propose here. However again, this is used to post-process the output of ABC, rather than to help choose the summary statistics. 
We look at differences between our approach and that of \cite{Beaumont:2002} empirically in the examples.

\subsection{Why use ABC?} \label{S:WHY}

%WHY USE ABC? DISCUSS RESULTS ON CALIBRATION/ADV AND MAIN THEOREM.

Our approach involves using simulation to find estimates of the posterior mean of each parameter. A natural question is why not use these estimates directly? We think using ABC has two important advantages over just using these estimates directly. The first is that ABC gives you a posterior distribution, and thus you can quantify uncertainty in the parameters as well as getting point estimates.

Moreover we have the following result.
\begin{theorem} \label{thm:4}
Let $\tilde{\theta} = \mbox{E}(\theta | S(\yobs))$.  Then for any function $g$,
\[
\mbox{E}\left(L(\theta, \tilde{\theta}; A) | S(\yobs)\right) \leq
\mbox{E}\left(L(\theta, g(S(\yobs)); A) | S(\yobs)\right).
\]
Furthermore, asymptotically as $h\rightarrow0$ the ABC posterior mean estimate of $\theta$ is optimal amongst estimates based on $S(\yobs)$.
\end{theorem}
{\bf{Proof:}}
The proof of the first part is the standard argument that the mean is the optimal estimator under quadratic loss \cite{Bernardo/Smith:1994}.
%That is, make the decomposition
%\begin{eqnarray*}
%\mbox{E} \left(L(\theta, g(S(\yobs)); A) | S(\yobs) \right) =
%\mbox{E} \left((\theta - \tilde{\theta})^T A (\theta - \tilde{\theta}) | S(\yobs)\right) \\
%+ \left(\tilde{\theta} - g(S(\yobs))\right)^T A \left(\tilde{\theta} - g(S(\yobs))\right),
%\end{eqnarray*}
%and, recalling that $A$ is positive-definite, the result holds by the non-negativity of the second term. 
The second part follows because as $h\rightarrow0$ the ABC posterior mean tends to $\tilde{\theta}$.
\hfill $\Box$

Note that this result states that, in the limit as $h\rightarrow 0$, ABC gives estimates that are at least as or more accurate than any other estimators based on the same summary statistics. 

{\bf Comparison with Indirect Inference}

Indirect inference \cite[]{Gourieroux:1993} is a method similar to ABC in that it uses simulation from a model to produce estimates of the model's parameters. The general procedure involves first analysing the data under an approximating model, and estimating the parameters, called auxilary parameters, for this model. Then data is simulated for a range of parameter values, and for each simulated data set we get an estimate of the auxilary parameters. Finally we estimate the true parameters based on which parameter values produced estimates of the auxilary parameters closest to those estimated from the true data. (In practice we simulate multiple data sets for each parameter value, and get an estimate of the auxilary parameters based on these multiple data sets.) The link to ABC is that the auxilary parameters in indirect inference are equivalent to the summary statistics in ABC. Both methods then use (different) simulation approaches to produce estimates of the true parameters from the values of the auxilary paramaters (or summary statistics) for the real data. 

For many approximating models, the auxilary parameters depend on a small set of summary statistics of the data; these are called auxilary statistics in \cite{Frigessi:2004}. In these cases indirect inference is performing inference based on these auxilary statistics. The above result shows that in the limit as $h\rightarrow 0$, ABC will be more accurate than an indirect inference method whose auxilary statistics are the same as the summary statistic used for ABC. We investigate this empirically in Section \ref{S:Ex}. %Note that \cite{Drovandi/Pettitt/Faddy:2011} also consider the links between indirect inference and ABC, but they focus on using the idea of an auxillary model to construct the summary statistics to be used within ABC.

\section{Examples} \label{S:Ex}

The performance of semi-automatic ABC was investigated in a range of examples:
independent draws from a complex distribution (Section \ref{S:Ex1}),
a stochastic kinetic network for biochemical reactions (Section \ref{S:ExSKN}),
a partially observed M/G/1 queue (Section \ref{S:ExMG1}),
an ecological population size model (Section \ref{S:ExRicker}),
and a Tuberculosis transmission model (Section \ref{S:Ex2}).
Section \ref{S:ExImpl} describes implementation details that are common to all examples.  Section \ref{S:ExSKN} concerns a dataset for which ABC and related methods have not previously been used, and highlights the use of noisy ABC in the sequential approach of Section \ref{S:mult}.  The other examples have previous analyses in the literature, and we show  that semi-automatic ABC compares favourably against existing methods including indirect inference, the synthetic likelihood method of \cite{Wood:2010} and ABC with ad-hoc summary statistics, with or without the regression correction method of \cite{Beaumont:2002}. We also show that direct use of the linear predictors created during semi-automatic ABC can be inaccurate (e.g Section \ref{S:Ex2}).
Outside Section \ref{S:ExSKN} noisy ABC runs are not shown; they are similar to non-noisy semi-automatic ABC but slightly less accurate. %\cite[]{Prangle:2011}.
The practical details of implementing our method are also explored, in particular how the choice of explanatory variables $f(\y)$ is made.

\subsection{Implementation details} \label{S:ExImpl}

Apart from the sequential implementation of ABC in \ref{S:ExSKN}, all ABC analyses  were performed using Algorithm \ref{Alg:3} with a Normal transition kernel.  The density kernel was uniform on an ellipsoid $\x^T A \x <c$.  This is a common choice in the literature and close to optimal for runs using semi-automatic ABC summary statistics as discussed in Section \ref{S:Links}.  For summary statistics not generated by our method we generally we used $A=I$. %Because all on a similar scale
In Section \ref{S:ExRicker} a different choice was necessary and is discussed there.
For ABC using summary statistics from our method, recall from Section \ref{sec:SS} that $A$ defines the relative weighting of the parameters in our loss function.
In Sections \ref{S:Ex1} and \ref{S:Ex2} the parameters are on similar scales so we used $A=I$.
Elsewhere, marginal parameter variances were calculated for the output of each pilot run and the means of these $(s_1^2, s_2^2, \ldots)$ taken. A diagonal $A$ matrix was formed with $i$th diagonal entry $s_i^{-2}$.

Other tuning details required by Algorithm \ref{Alg:3} are the choice of $h$, the variance matrix of the transition kernel and the starting values of the chain.  Where possible, these were chosen by manual experimentation or based on previous analyses (e.g.~from pilot runs).  Otherwise they were based on a very short ABC rejection sampling analysis.  Except where noted otherwise, $h$ was tuned to give an acceptance rate of roughly $1$\% as this gave reasonable results in the applications considered. \cite[An alternative would be to use computational methods that try and choose $h$ for each run, see][]{Bortot:2007,Ratmann:2007}.

In the following examples our method is compared against an ABC analysis with summary statistic based on the existing literature, referred to as the ``comparison'' analysis.
To allow a fair comparison, this uses the same number of simulations as the entire semi-automatic method and a lower acceptance rate, roughly $0.5$\%.
%Note that it was sometimes necessary to modify the CPU time division discussed in Section \ref{S:Alg} reducing the CPU time on step (ii) due to limitations on available memory.
%Also, explanatory variables simulated as training data were sometimes collinear.  When this occured, variables with unidentifiable coefficients were discarded.

For simulation studies on multiple datasets, the accuracies of the various analyses were compared as follows. The point estimate for each dataset was calculated, and the quadratic loss (\ref{eq:loss}) of each parameter estimate relative the true parameter value was calculated.
%The $A$ matrix used in this calculation is that described above for use in the semi-automatic ABC acceptance kernel.
%The mean loss over all datasets is then presented,
We present the mean quadratic losses of the individual parameters. In tables of results we highlight the smaller quadratic losses (all with $10\%$ of the smallest values) by italicising.

%Semi-automatic ABC only makes a small number of summary statistics available for regression correction.  We tried using all the explanatory variables in regression correction.  This sometimes improves results but often produces wildly inaccurate predictions, as there is there is no guarantee that all of the simulated explanatory variables will be close to the observed values.

%Burn-in used where appropriate.
%Thinning typically used to save storage space.
%True pars sometimes used as initial state to avoid burn-in (e.g.~g and k).
%Normal MH steps used with variance tuned to be roughly equal to that of ABC posterior.

\subsection{Inference for $g$ and $k$ Distribution} \label{S:Ex1}

The $g$ and $k$ distribution is a flexibly shaped distribution used to model non-standard data through a small number of parameters \cite[]{Haynes:1998}.  It is defined by its inverse distribution function (\ref{eq:g&k}), below, but has no closed form density.  Likelihoods can be evaluated numerically but this is costly \cite[]{Rayner:2002, Drovandi/Pettitt:2009}. ABC methods are attractive because simulation is straightforward by the inversion method. Here we use the fact we can calculate the maximum likelihood estimate numerically to also compare ABC with a full-likelihood analysis.

The distribution is defined by:
\begin{equation} \label{eq:g&k}
 F^{-1} (x; A,B,c,g,k) = A + B \left( 1 + c \frac{1-\exp(-gz(x)}{1+\exp(-gz(x)} \right) (1+z(x)^2)^k z(x)
\end{equation}
where $z(x)$ is the $x$th standard Normal quantile,
$A$ and $B$ are location and scale parameters and $g$ and $k$ are related to skewness and kurtosis.
The final parameter $c$ is typically fixed as $0.8$, and this is assumed throughout, leaving unknown parameters $\theta = (A,B,g,k)$.
The only parameter restrictions are $B > 0$ and $k > -1/2$ \cite[]{Rayner:2002}.

\cite{Allingham:2008} used ABC to analyse a simulated dataset of $n=10^4$ independent draws from the $g$-and-$k$ distribution with parameters $\theta_0=(3,1,2,0.5)$.  A uniform prior on $[0,10]^4$ was used and the summary statistics were the full set of order statistics.  We studied multiple datasets of a similar form as detailed below.  Our aim is firstly to show how we can implement semi-automatic ABC in a situation where there are large numbers of possible explanatory variables (just using the order statistics gives $10^4$ explanatory variables), and to see how the accuracy of semi-automatic ABC compares with the use of arbitrarily chosen summary statistics in  \cite{Allingham:2008}. We also aim to look at comparing semi-automatic ABC with the linear regression correction of \cite{Beaumont:2002}, and with indirect inference.

{\bf Comparison of ABC methods}

The natural choice of explanatory variables for this problem are based upon the order statistics, and also powers of the order statistics. Considering up to the fourth power seems appropriate as informally the four parameters are linked to location, scale, skewness and kurtosis. However fitting the linear model with the resulting $4\times10^4$ exaplanatory variables is impracticable. As a result we considered using a subset of $m$ evenly spaced order statistics, together with up to $l$ powers of this subset. To choose appropriate values for $m$ and $l$ we fitted linear models with $m$ ranging over a grid of values between $60$ and $140$, and $l$ ranging between $1$ and $4$, and used BIC (averaged across the models for the four parameter values) to choose an appropriate value for $m$ and $l$. We then used the summary statistics obtained from the linear model with this value  of $m$ and $l$ in the final run of ABC. For simplicity we did this for the first data set, and kept the same value of $m$ and $l$  for analysing all subsequent data sets.

Note that using subsets of order statistics has computational advantages, as these can be generated efficiently by simulating corresponding standard uniform order statistics using the exponential spacings method of \cite{Ripley:1987} (page 98) and performing inversion by substituting these in (\ref{eq:g&k}).  The cost is linear in the number of order statistics required. Our pilot ABC run used the summary statistics from \cite{Allingham:2008}. Fitting the different linear models added little to the overall computational cost, which is dominated by the simulation of the data sets at the different stages of the procedure.

Our semi-automatic ABC procedure chose $m=100$ order statistics and $l=4$. The accuracy of the resulting parameter estimates, measured by square error loss across implementation of semi-automatic ABC on 50 data sets is shown in Table \ref{tab:1}. For comparison we show results of the \cite{Allingham:2008} ABC method, implemented to have the same overall computational cost. We also show the accuracy of estimates obtained by post-processing the \cite{Allingham:2008} results using the \cite{Beaumont:2002} regression correction. We could only use this regression correction on 48 of the 50 data sets, as for the remaining 2, there were too few acceptances in the ABC run for the regression correction to be stable (on those two data sets the resulting loss after performing the regression correction was orders of magnitude greater than the original ABC estimates).

Despite the \cite{Allingham:2008} analysis performing poorly, and hence producing a poor pilot region for semi-automatic ABC; semi-automatic ABC appears to perform well. It has losses that are between a factor of 2 and 100 smaller than the method of \cite{Allingham:2008} with or without the regression correction. Using the regression correction does improve accuracy of the estimates for three of the four parameters, but to a lesser extent than semi-automatic ABC.
For comparison, using the predictors from the linear-regression to directly estimate the parameters had similar accuracy for all parameters except $g$, where the linear-predictor's average error was greater by about a third.

To investigate the effect of the pilot run on semi-automatic ABC, and the choice of summary statistics on the regression correction, we repeated this analysis by implementing ABC with 100 order statistics. This greatly improved the performance of ABC, showing the importance of the choice of summary statistics. The improved pilot run also improves semi-automatic ABC but to a much lesser extent. In this case semi-automatic ABC has similar accuracy to the comparison ABC run with the regression correction. For a further comparison we also calculated the MLEs for each data set numerically. The two best ABC runs have mean quadratic loss that is almost identical to that of the MLEs.
 
\begin{table}[ht]
\begin{center}
\begin{tabular}{rcccc}
  \hline
  & A & B & g & k  \\ 
  \hline
  Allingham et al & 0.0059 & 0.0013 & 3.85 & 0.00063  \\ 
  Allingham + reg & 0.00040 & 0.0017 & 0.28 & 0.00051  \\
  Semi-Automatic ABC & {\em 0.00016} & {\em 0.00056}  & 0.044 & 0.00023 \\ \hline
  Comparison & 0.00025 & 0.00063 & 0.0061 & 0.00041\\
  Comparison + reg &{\em 0.00016} & {\em 0.00055} & {\em 0.0014} & {\em 0.00015} \\
  Semi-Automatic ABC &{\em 0.00015} & {\em 0.00053}& {\em 0.0014} & {\em 0.00015} \\ \hline
   MLE & 0.00016 & 0.00055 & 0.0013 & 0.00014  \\
  \hline
\end{tabular}
\end{center}
\caption{\footnotesize{Mean quadratic losses of various ABC analyses of 50 $g$-and-$k$ datasets with parameters $(3,1,2,0.5)$. The first three rows are based on using the Allingham et al. summary statistics in ABC, and in the ABC pilot run for semi-automatic ABC. The next three rows use just 100 evenly spaced order statistics. Results based on using the \cite{Beaumont:2002} regression correction are denoted {\em reg}. For Allingham + reg, we give the mean loss for just 48 of the 50 data sets. For the remaining 2 data sets the number of ABC acceptances was low $\approx 200$, and the regression correction was unstable. For comparison we give the mean quadratic loss of the true MLEs.}}
\label{tab:1}
\end{table}

{\bf Comparison with Indirect Inference}

Theorem \ref{thm:4} shows that asymptotically ABC is at least as accurate as other estimators based on the same summary statistics.
We tested this with a comparison against indirect inference (see Section \ref{S:WHY}). %and \cite{Frigessi:2004}
The  semi-automatic ABC analysis was repeated, and to give a direct comparison, its summary statistics were used as the indirect inference auxiliary statistics.

\begin{figure}[ht] \begin{center}
  \includegraphics[totalheight=4in]{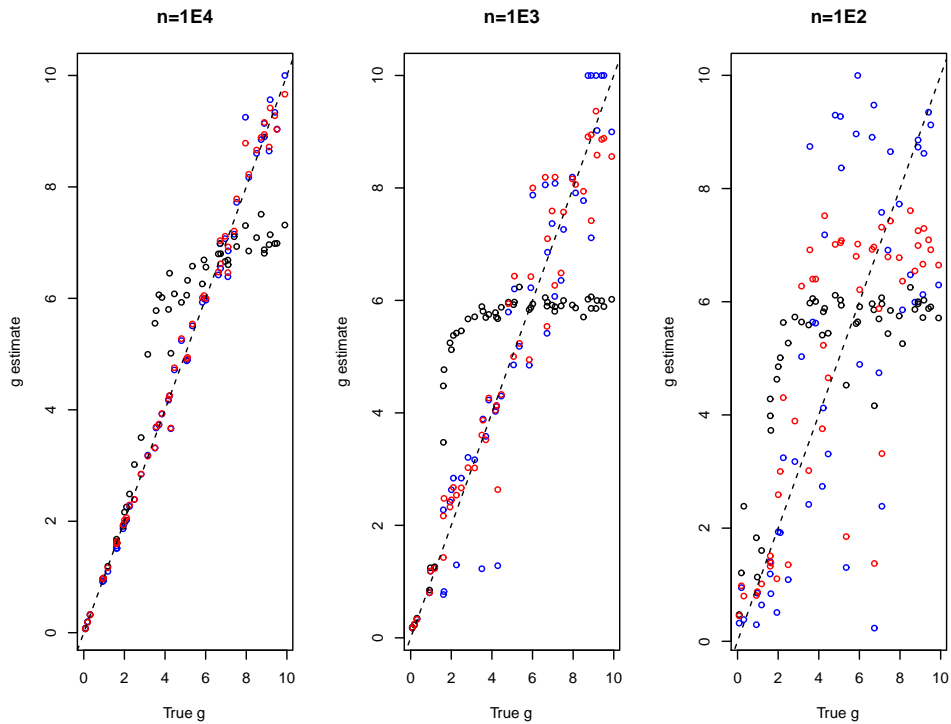}
  \caption{\footnotesize{Estimated $g$ values from indirect inference (blue) and semi-automatic ABC pilot (black) and final (red) analyses, plotted against true $g$ values for 50 $g$-and-$k$ datasets of sample size $n$.}}
  \label{fig:ii}
\end{center} \end{figure}

 %Indirect inference was implemented by a numerical optimisation routine \cite[]{}.
Initial analysis showed that which method is more accurate depends on the true value of $\theta$ and in particular the parameter $g$; this is illustrated by Figure \ref{fig:ii}.
Therefore we studied datasets produced from variables $g$ values;
we drew 50 $g$ values from its prior %stratified to reduce variance.
and for each simulated datasets of $n$ $g$-and-$k$ draws conditional on $\theta=(3,1,g,0.5)$ for $n=10^2, 10^3$ and $10^4$.

Each semi-automatic ABC analysis used a total of $3.1 \times 10^6$ simulated data sets.  Indirect inference was roughly tuned so that the total number of simulations equalled this. (Similar results can be obtained from indirect inference using many fewer simulations, and indirect inference is thus a computationally quicker algorithm).  Mean losses are given in Table \ref{tab:3}, showing that while the methods perform similarly for $n=10^4$, ABC is more accurate for smaller $n$.

More detail is given in Figure \ref{fig:ii} which plots the true against estimated $g$ values.
Of particular interest is the $n=100$ case where the $g$ parameter is very hard to identify for $g>3$.
It is over this range that ABC out-performs indirect inference most clearly, with estimates from indirect inference being substantially more variable than those for ABC.

\begin{table}[ht]
\begin{center}
\begin{tabular}{crcccc}
  \hline
  && A & B & g & k\\ 
  \hline
  & Pilot & 0.0003 & 0.0008 & 1.7 & 0.0004  \\ 
  $n=10^4$ & Indirect inference & 0.0003 & 0.0022 & 0.082 & 0.0063  \\ 
  & Semi-automatic ABC & {\em 0.0001} & {\em 0.0005} & {\em 0.059} & {\em 0.0002}  \\ 
  \hline
  & Pilot & 0.0031 & 0.014 & 4.6 & 0.0073  \\ 
  $n=10^3$ & Indirect inference & 0.0066 & 0.014 & 0.83 & 0.0053  \\ 
  & Semi-automatic ABC & {\em 0.0012} & {\em0.0094} & {\em0.51} & {\em0.0042}  \\ 
  \hline
  & Pilot & 0.0089 & 0.039 & 4.8 & 0.057 \\ 
  $n=10^2$ & Indirect inference & 0.018 & 0.059 & 5.5 & 0.067 \\ 
  & Semi-automatic ABC & {\em0.0075} & {\em0.046} & {\em3.5} & {\em0.040}  \\ 
  \hline
\end{tabular}
\end{center}
\caption{\footnotesize{Mean quadratic losses of semi-automatic ABC and indirect inference analyses of 50 $g$-and-$k$ datasets with variable $g$ parameters.}}
\label{tab:3}
\end{table}

\subsection{Inference for Stochastic Kinetic Networks} \label{S:ExSKN}

Stochastic kinetic networks are used to model biochemical networks. The state of the network is determined by the number of each of a discrete set of molecules, and evolves stochastically through reactions between these molecules. See \cite{Wilkinson:2009} and references therein for further background.

Inference for these models is challenging, as the transition density of the model is intractable. However simulation from the models is possible, for example using the Gillespie Algorithm \cite[]{Gillespie:1977}. As such they are natural applications for ABC methods. Here we focus on a simple example of a stochastic kinetic network, the Lotka-Volterra model of \cite{Boys/Wilkinson/Kirkwood:2008}. While simple, \cite{Boys/Wilkinson/Kirkwood:2008} show the difficulty of full-likelihood inference.

This model has a state which consists of two molecules. Denote the state at time $t$ by $\Y_t=(Y_t^{(1)},Y_t^{(2)})$. There are three types of reaction: the birth of a molecule of type 1, the death of a molecule of type 2, and a reaction between the two molecules which removes a type 1 molecule and adds a type 2 molecule. (The network is called the Lotka-Volterra model, due to its link with predator-prey models, type 1 molecules being prey and type 2 being predators.)

The dynamics for the model are Markov, and can be specified in terms of transition over a small time-interval $\delta t$. For positive parameters $\theta_1$, $\theta_2$ and $\theta_3$, we have
\begin{eqnarray*}
\lefteqn{\Pr(\Y_{t+\delta t}=(z_1,z_2) | \Y_t=(y_1,y_2) )} & & \\ &=& \left\{ \begin{array}{cl}
1-(\theta_1y_1+\theta_2y_1y_2+\theta_3y_2)\delta t + o(\delta t) & \mbox{if $z_1=y_1$ and $z_2=y_2$}, \\
\theta_1y_1\delta t + o(\delta t) & \mbox{if $z_1=y_1+1$ and $z_2=y_2$}, \\
\theta_2y_1y_2\delta t + o(\delta t) & \mbox{if $z_1=y_1-1$ and $z_2=y_2+1$}, \\
\theta_3 y_3\delta t + o(\delta t) & \mbox{if $z_1=y_1$ and $z_2=y_2-1$,} \\
o(\delta t) & \mbox{otherwise.} 
\end{array} \right.
\end{eqnarray*}
We will focus on the case of the network being fully observed at discrete time-points, and also on just observing the type 2 molecules initially, together with the type 1 molecules at all observation points. All simulations use the parameters from \cite{Boys/Wilkinson/Kirkwood:2008}, with $\theta_1=0.5$, $\theta_2=0.0025$, and $\theta_3=0.3$, and evenly sampled data collected at time-intervals of length $\tau$. We will perform analysis conditional on the known initial state of the system.

{\bf Sequential ABC Analysis}

\cite{Wilkinson:2011} consider simulation-based approaches for analysing stochastic kinetic networks which are based upon sequential Monte Carlo methods \cite[see][for an introduction]{Doucet/Godsill/Andrieu:2000}. In some applications, to get these methods to work for reasonable computational cost, \cite{Wilkinson:2011} suggests using ABC. A version of such an algorithm (though based on importance sampling rather than MCMC for the sequential update) for the Lotka-Volterra model is given in Algorithm \ref{Alg:4}. Note there are many approaches to improve the computational efficiency of this algorithm see for example \cite{Doucet/Godsill/Andrieu:2000} and \cite{Doucet/DeFreitas/Gordon:2001} for details.
\begin{algorithm}[ht]
\begin{footnotesize}
\rule{\textwidth}{1mm}
\underline{Sequential ABC for Lotka-Volterra Model}
\\
\begin{tabular}{ll}
{\bf Input:} &A set of times, $t_0,\ldots,t_n$, and data values, $\y_{t_0},\ldots,\y_{t_n}$. \\
& A number of particles, $N$, a kernel, $K(\cdot)$, and a bandwidth, $h$. \\
\\
{\bf Initialise:} &For $i=1,\ldots,N$ sample $\theta^{(i)}=(\theta^{(i)}_1,\theta^{(i)}_2,\theta^{(i)}_3)$ from the prior distribution for the parameters.
\\
\\ 
\end{tabular}
{\bf Iterate:} For $j=1,2,\ldots,n$
\begin{enumerate}
\item For $i=1,\ldots,N$, sample a value for the state at time $t_j$, $\y^{(i)}_{t_j}$,  given its value at time $t_{j-1}$, $\y_{t_{j-1}}$, and $\theta^{(i)}$ using the Gillespie algorithm.
\item For $i=1,\ldots,N$, calculate weights
\[
w^{(i)}=K\left( [\y^{(i)}_{t_j}-\y_{t_j}]/h \right)
\]
\item Sample $N$ times from a Kernel density approximation to weighted sample of $\theta$ values, $\{\theta^{(i)},w_i\}_{i=1}^N$ \cite[see e.g.][]{Liu/West:2001}. Denote this sample $\theta^{(1)},\ldots,\theta^{(N)}$.
\end{enumerate}
{\bf Output:} A sample of $\theta$ values. \\
\rule{\textwidth}{1mm}
\caption{\footnotesize A sequential ABC sampler for the Lotka-Volterra model. \label{Alg:4}}
\end{footnotesize}
\end{algorithm}

The discussion following Theorem \ref{thm:5} shows that an algorithm like Algorithm \ref{Alg:4} may give inconsistent parameter estimates, even when ignoring the Monte Carlo error. A simple remedy to this is to implement noisy ABC within this algorithm. This can be done by adding noise to the observed values. The noisy sequential ABC algorithm is given in Algorithm \ref{Alg:5}.

\begin{algorithm}[ht]
\begin{footnotesize}
\rule{\textwidth}{1mm}
\underline{Noisy Sequential ABC}\\
Replace step 2 in Algorithm \ref{Alg:4} with: \\
\begin{enumerate}
\item[2.] Simulate $\x_j$ from $K(\x)$. For $i=1,\ldots,N$, calculate weights
\[
w^{(i)}=K\left( [\y^{(i)}_{t_j}-\y_{t_j}-h\x_j]/h \right)
\]
\end{enumerate}
\rule{\textwidth}{1mm}
\caption{\footnotesize Change of Algorithm \ref{Alg:4} for noisy sequential ABC. \label{Alg:5}}
\end{footnotesize}
\end{algorithm}

To evaluate the relative merits of the two sequential ABC algorithms, we analysed 100 simulated data sets for the Lotka-Volterra model. For stability of the sequential Monte Carlo algorithm we chose $K(\cdot)$ to be the density function of a bivariate normal random variable, as a uniform kernel can lead to iterations where all weights are 0. We analysed two data scenarios, with $\tau=0.1$, one with full observations, and one where only the number of type 1 molecules is observed. The sequential ABC algorithms were implemented with $N=5,000$ and $h=\sqrt{\tau}$. The latter was chosen to be a small value for which the sequential algorithms still performed adequately in terms of Monte Carlo performance (as measured by variability of the weights after each iteration).

\begin{figure}[ht] \begin{center}
  \includegraphics[totalheight=4in]{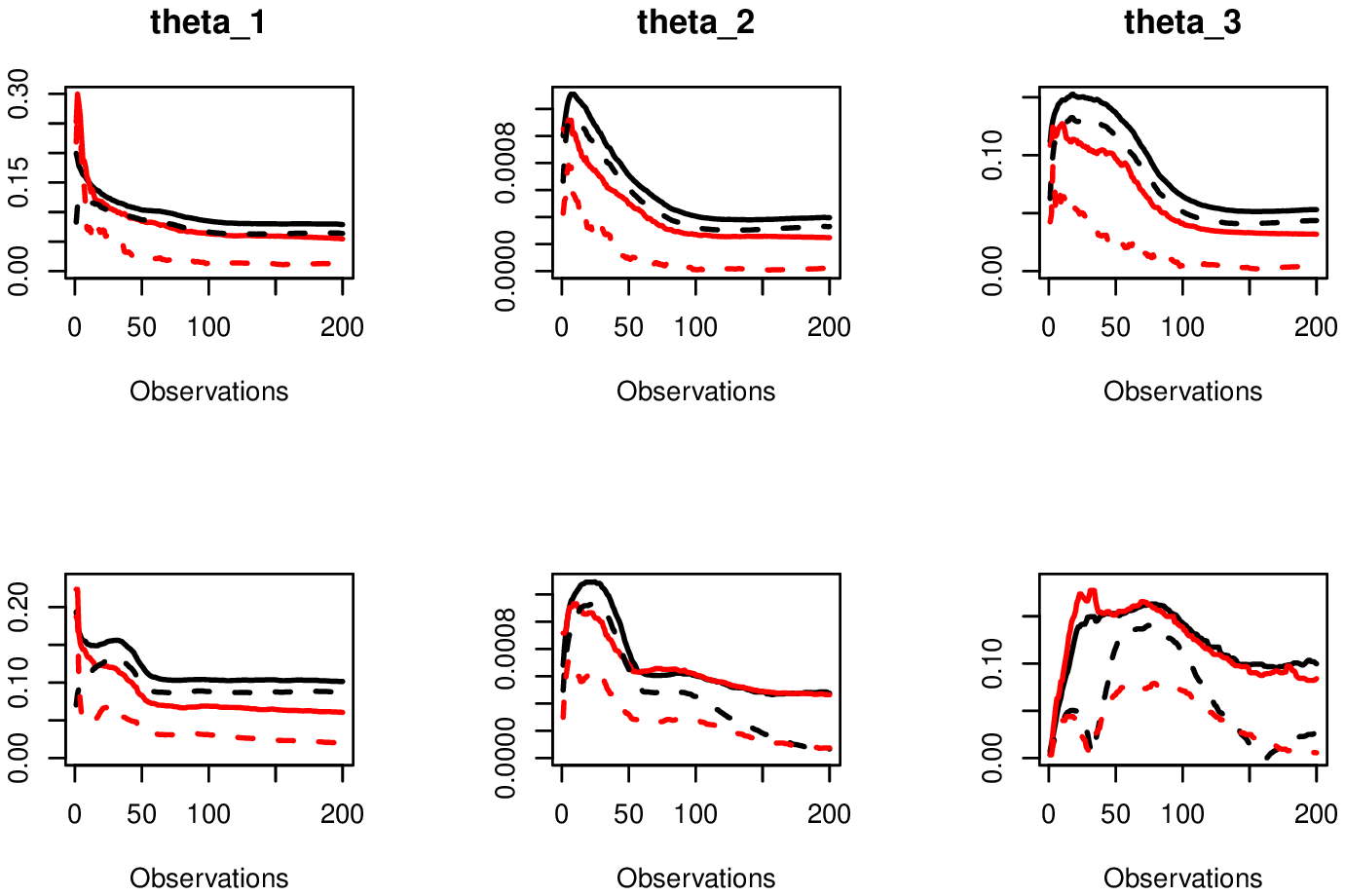}
  \caption{\footnotesize{Plots of root square error loss (full-lines) and absolute bias (dashed lines) for the standard sequential ABC algorithm (black), and the noisy ABC version (red) as a function of the number of observations. Top row is observing the number of both molecules; bottom row is for only observing the number of type 1 molecules. Both have $\tau=0.1$.}}
  \label{fig:LV}
\end{center} \end{figure}

Results are shown in Figure \ref{fig:LV}. These show both the absolute bias and the root mean quadratic loss for estimating each parameter for both ABC algorithms as the number of observations analysed varies between 1 and 200. For the full observation case, we can see evidence of bias in all three parameters for the standard sequential ABC algorithm. Noisy ABC shows evidence of being asymptotically unbiased as the number of observations increases. Overall, noisy ABC appears more accurate, except perhaps if only a handful of observations are made. When just type 1 molecules are observed, the picture is slightly more complex. We observed evidence of bias for the standard ABC algorithm for $\theta_1$, and for this parameter noisy ABC is more accurate. For the other parameters, there appears little difference between the accuracy and bias of the two ABC algorithms. The difference between the parameters is likely to be because $\theta_1$ only affects the dynamics of the observed molecule.

\subsection{Inference for Ricker model} \label{S:ExRicker}

The Ricker map is an example of an ecological model with complex dynamics. It updates a population size $N_t$ over a time step by
\[
N_{t+1} = r N_t e^{-N_t + e_t}
\]
where the $e_t$ are independent $N(0,\sigma^2_e)$ noise terms. \cite{Wood:2010} studies a model in which Poisson observations $y_t$ are made with means $\phi N_t$. The parameters of interest are $\theta = (\log r, \sigma_e, \phi)$. The initial state is $N_0 = 1$ and the observations are $y_{51}, y_{52}, \ldots, y_{100}$. This is a complex inference scenario in which there is no obvious natural choice of summary statistics. Furthermore \cite{Wood:2010} argues that estimating parameters by maximum likelihood is difficult for this model due to the chaotic nature of the system.

\cite{Wood:2010} analyses this model using a Metropolis-Hastings algorithm to explore a ``synthetic likelihood'', $L_s(\theta)$. A summary statistic function must be specified as in ABC, and $L_s(\theta)$ is then the density of $\Sobs$ under a $N(\mu_\theta, \Sigma_\theta)$ model where $\mu_\theta$ and $\Sigma_\theta$ are an approximation of the mean and variance of $\Sobs$ for the given parameter value obtained through simulation \cite[see][for more details]{Wood:2010}. %the mean and variance of 500 summary statistic vectors simulated conditional on $\theta$. % as follows. First note that a dataset can be represented as $g(\theta, e, u)$ where $e$ is a vector of 100 standard normal draws (to produce the $e_t$ terms) and $u$ is a vector of 100 standard uniform draws (to generate the Poisson draws). Initially vectors $e^1, e^2, \ldots, e^{500}$ and $u^1, u^2, \ldots, u^{500}$ are simulated, and the datasets $g(\theta, e^1, u^1), \ldots, g(\theta, e^{500}, u^{500})$ are used to calculate $L_s(\theta)$.

We simulated 50 datasets with $\log r = 3.8$, $\phi = 10$ and $\log \sigma_e$ values drawn from a uniform distribution on $[\log 0.1, 0]$ (Initial analysis showed that the true $\log \sigma_e$ value affected the performance of the different methods).
%This is because estimates of this value are typically biased (downwards) so for methods with more weakly identified results the prior has a large influence in improved or worsening the estimate, and we need to average out these effects.
These were analysed under our approach and that of \cite{Wood:2010} (using the code in that paper's supplementary material).
The distribution just mentioned was used as prior for $\log \sigma_e$, and improper uniforms priors were placed on $\log r \geq 0$ and $\phi$.  All parameters were assumed independent under the prior.
As each parameter had a non-negativity constraint, the MCMC algorithms used log transforms of $\theta$ as the state.
Each semi-automatic ABC analysis used $10^6$ simulated datasets.

The summary statistics used by \cite{Wood:2010} were
the autocovariances to lag 5, %calculated by SW's code
the coefficients of a cubic regression of the ordered differences $\Delta_t = y_t - y_{t-1}$ on those of the observed data,
least squares estimates for the model
$y_{t+1}^{0.3} = \beta_1 y_t^{0.3} + \beta_2 y_t^{0.6} + \epsilon_t$,
the mean observation, $\bar{y}$,
and $\sum_{t=51}^{100} \mathbb{I}(y_t = 0)$ (the number of zero observations).
We denote this set $E_0$ and use it in the semi-automatic ABC pilot runs.
The pilot runs used acceptance kernel $A=\Sigma^{-1}$ where $\Sigma$ is the sample variance matrix of 500 simulated summary statistics vectors for a representative fixed parameter value.

In the regression stage of semi-automatic ABC, training datasets mostly comprised of zeros were fitted poorly.  Since the observed datasets had at most 31 zeros, any datasets with 45 or more zeros was discarded from our training data, and automatically rejected in the subsequent ABC runs. %before the main summary statistics were calculated.
Summary statistics were constructed for two nested sets of explanatory variables.
The smaller set, E1, included E0 and additionally
$\sum_{t=51}^{100} \mathbb{I}(y_t = j)$ for $1 \leq j \leq 4$,
$\log \bar{y}$,
log of the sample variance,
$\log \sum_{i=51}^{100} y_i^j$ for $2 \leq j \leq 6$ and
autocorrelations up to lag 5.
The larger set, E2, also  added
$(y_t)_{51 \leq t \leq 100}$ (time ordered observations),
$(y_{(t)})_{1 \leq t \leq 50}$ (magnitude ordered observations),
$(y_t^2)_{51 \leq t \leq 100}$,
$(y_{(t)}^2)_{1 \leq t \leq 50}$,
$(\log(1+y_t))_{51 \leq t \leq 100}$,
$(\log(1+y_{(t)}))_{1 \leq t \leq 50}$,
$(\Delta_t^2)_{52 \leq t \leq 100}$ and
$(\Delta_{(t)}^2)_{1 \leq t \leq 49}$.

Using E2 instead of E1  reduced BIC in each linear regression by the order of thousands for all but one dataset, suggesting that E2 gives better predictors. Thus we used the summary statistics based on E2 within the ABC analysis. %, which is borne out by the results of Table \ref{tab:Ricker}.
%While the E1 analyses perform slightly worse than synthetic likelihood and the comparison ABC analysis using E0 as summary statistics, 
The results for these ABC analyses show an improvement over the synthetic likelihood for estimating $\log r$ and $\phi$, and identical performance for estimating $\sigma_{e}$. The semi-automatic ABC analysis also does better than the comparison ABC one (based on summary statistics $E_0$). Note that for this application the linear regression adjustment of \cite{Beaumont:2002} actually produces worse results then using the raw output of the comparison ABC analyses.

Finally we looked at 95\% credible intervals that were constructed from the semi-automatic ABC and synthetic likelihood method.
The coverage frequencies of these intervals were $0.86$, $0.70$, $0.96$ for synthetic likelihood and $0.98$, $0.92$, $1$ for our method. Whilst the synthetic likelihood intervals appear to have coverage frequencies that are too low for two of the parameters, those from ABC are consistent with $0.95$ coverage given a sample size of 50 datasets.

% latex table generated in R 2.11.0 by xtable 1.5-6 package
% Thu Feb 17 09:47:29 2011
\begin{table}[ht]
\begin{center}
\begin{tabular}{rcccc}
  \hline
  & $\log r$ & $\sigma_e$ & $\phi$  \\ 
  \hline
  Synthetic likelihood & 0.050 & {\em0.032} & 0.66 \\ 
  Comparison & {\em 0.039} & 0.038 & 0.54 \\
  Comparison + regression & 0.046 & 0.041 & 0.78  \\  
  %Semi-automatic ABC E1 & 0.028 & 0.039 & 0.80  \\ 
  Semi-automatic ABC  & {\em 0.039} & {\em 0.032} & {\em 0.36}  \\ 
 % Linear predictors E1 & 0.041 & 0.043 & 1.12  \\ 
 % Linear predictors  & {\em 0.036} & {\em 0.031} & {\em 0.39}  \\ 
  \hline
\end{tabular}
\end{center}
\label{tab:Ricker}
\caption{\footnotesize{Mean quadratic losses for various analyses of 50 simulated Ricker datasets.}}
\end{table}
\subsection{Inference for M/G/1 Queue} \label{S:ExMG1}

Queuing models are an example of stochastic models which are easy to simulate from, but often have intractable likelihoods. It has been suggested to analyse such models using simulation-based procedures, and we will look at a specific M/G/1 queue that has been analysed by both ABC \cite[]{Blum/Francois:2010} and indirect inference \cite[]{Frigessi:2004} before. 
In this model, the service times uniformly distributed in the interval $[\theta_1, \theta_2]$ and inter-arrival times exponentially distributed with rate $\theta_3$.  The queue is initially empty and only the inter-departure times $y_1, y_2, \ldots, y_{50}$ are observed.  

We analysed 50 simulated datasets from this model. The true parameters were drawn from the prior under which $(\theta_1, \theta_2 - \theta_1, \theta_3)$ are uniformly distributed on $[0,10]^2 \times [0,1/3]$.
%In previous analyses $\theta_3$ prior had support [0,10]
%Frigessi used fixed parameter values 0.3, 0.9, 1 and the data consisted of 100 interdeparture times from ``the stable queue'' (50 data sets).
%Blum used fixed parameter values 1, 5, 0.2 and 50 observations (1 data set).
This choice gives arrival and service times of similar magnitudes, avoiding the less interesting situation where all $y_i$ values are independent draws from a single distribution.

The analysis of \cite{Blum/Francois:2010} used as summary statistics evenly spaced quantiles of the interdeparture times, including the minimum and maximum.
Our semi-automatic ABC pilot analyses replicate this choice, using 20 quantiles.
The explanatory variables $f(\y)$ we used to construct summary statistics were the ordered interdeparture times.  Adding powers of these values to $f(\y)$ produced only minor improvements so the results are not reported.  The analysis of each dataset used $10^7$ simulated datasets, split in the usual way.

We also applied the indirect inference approach of \cite{Frigessi:2004}.  This used auxiliary statistics $(\bar{y}, \min y_i, \hat{\theta}_2^{ML})$ where %$\bar{y}$ is the mean interdeparture time and
$\hat{\theta}_2^{ML}$ is the maximum likelihood estimate of $\theta_2$ under an auxiliary model which has a closed form likelihood; independent observations from the steady state of the queue. Numerical calculation of $\hat{\theta}_2^{ML}$ is expensive so indirect inference used many fewer simulated datasets than ABC but had similar runtimes.

Table \ref{tab:MG1} shows the results.
Semi-automatic ABC outperforms a comparison analysis using 20 quantiles as the summary statistics, but once a regression correction is applied to the latter the results become very similar. 
%The results illustrate that a near automatic application of our approach can be effective in a relatively simple setting.
Note that here the semi-automatic linear predictors are less accurate when used directly rather than in ABC. Indirect inference is more accurate at estimating $\theta_2$, presumably due to the accuracy of $\hat{\theta}_2^{ML}$ as an estimate of $\theta_2$. However it is still substantially less accurate for the other two parameters. One advantage of indirect inference, is that as it requires less simulations to estimate the parameters accurately, it can more easily accommodate summaries that are expensive to calculate, such as $\hat{\theta}_2^{ML}$.

% latex table generated in R 2.12.1 by xtable 1.5-6 package
% Mon Feb 14 11:45:31 2011
\begin{table}[ht]
\begin{center}
\begin{tabular}{rcccc}
  \hline
 & $\theta_1$ & $\theta_2$ & $\theta_3$ \\ 
  \hline
  %Pilot & 1.1919 & 2.1745 & 0.0013 & 0.4583 \\ 
  %Pilot + regression & 0.0206 & 1.0645 & 0.0013 & 0.1949 \\ 
  Comparison & 1.1 & 2.2 & 0.0013  \\ 
  Comparison + regression & {\em 0.020} &  1.1 & {\em 0.0013} \\ 
  Semi-automatic ABC & {\em 0.022} & 1.0 & {\em 0.0013}  \\ 
%  Semi-automatic ABC + regression & 0.0225 & 0.97 & 0.0013  \\ 
  %Main 3 powers & 0.0257 & 1.0007 & 0.0012 & 0.1881 \\ 
  %Main 3 powers + sumstat regression & 0.0242 & 0.9328 & 0.0012 & 0.1827 \\ 
  Semi-automatic predictors & 0.024 & 1.2 & 0.0017 \\ 
  %Raw3 & 0.0241 & 1.0408 & 0.0013 & 1.0663 & 0.1973 \\ 
  Indirect inference & 0.17 & {\em 0.46} & 0.0035 \\
   \hline
\end{tabular}
\end{center}
\caption{\footnotesize{Mean quadratic losses for various analyses of 50 M/G/1 datasets.}}
\label{tab:MG1}
\end{table}

\subsection{Inference of Tuberculosis Transmission} \label{S:Ex2}

\cite{Tanaka:2006} use ABC to analyse Tuberculosis bacteria genotype data sampled in San Francisco over a period from 1991 to 1992.  Table \ref{tab:4} shows the data, consisting of 473 bacteria samples split into clusters which share the same genotype on a particular genetic marker. Thus the data consists of 282 bacteria samples that had unique genotypes, 20 pairs of bacteria that had the same genotype, and so on.

\begin{table}[htp]
\begin{center}
\begin{tabular}{l|cccccccccc}
Cluster size      & 1  &2  &3  &4 &5 &8 &10 &15 &23 &30 \\
\hline
Number of clusters&282 &20 &13 &4 &2 &1 &1  &1  &1  &1 \\
\end{tabular}
\end{center}
\caption{\footnotesize{Tuberculosis bacteria genotype data.}}
\label{tab:4}
\end{table}
The proposed model was based on an underlying continuous time Markov process.
Denote the total number of cases at time $t$ by $N(t)$.
The process starts at $t=0$ with $N(0)=1$.
There are three types of event: birth, death (encompassing recovery of the host) and mutation.
The rate of each event type is the product of $N(t)$ and the appropriate parameter:
$\alpha$ for birth, $\delta$ for death and $\theta$ for mutation. It was assumed that each mutation creates a completely new genotype.
Cluster data is a simple random sample of 473 cases taken at the first $t$ such that $N(t)=10,000$.
The model conditions on such a $t$ existing in the underlying process.

%Reparameterisation
This data contains no information on time, so for $k>0$, parameter values $(\alpha, \delta, \theta)$ and $(k\alpha, k\delta, k\theta)$ give the same likelihood.
We reparameterise to $(a,d,\theta)$ where $a = \alpha / (\alpha+\delta+\theta)$ and $d = \delta / (\alpha+\delta+\theta)$.
The likelihood under this parameterisation depends only on $a$ and $d$.
%Choice of prior
To reflect prior ignorance of $(a,d)$ we use the prior density $\pi(a,d,\theta) \propto \pi(\theta) \mathbb{I}(0 \leq d \leq a) \mathbb{I}(a+d < 1)$, where $\pi(\theta)$ is the marginal prior for $\theta$ used in \cite{Tanaka:2006}.
The prior restriction $d \leq a$ avoids the need for simulations in which $N(t)=10,000$ is highly unlikely to occur.
The other prior restrictions follow from positivity constraints on the original parameters.
Under this prior and parameterisation, the marginal posterior of $\theta$ is equal to its prior and the problem reduces to inference on $a$ and $d$.

%Previous analyses (in particular: summary statistics)
\cite{Tanaka:2006} used two summary statistics for their ABC analysis: $g/473$ and $H = 1 - \sum_i (n_i / 473)^2$, where $g$ is the number of distinct clusters in the sample, and $n_i$ is the number of observed samples in the $i$th genotype cluster.  We retain this choice for our semi-automatic ABC pilot and the comparison ABC analysis reported below.
%Pilot details

\begin{figure}[htp] \begin{center}
  \includegraphics[totalheight=3.2in]{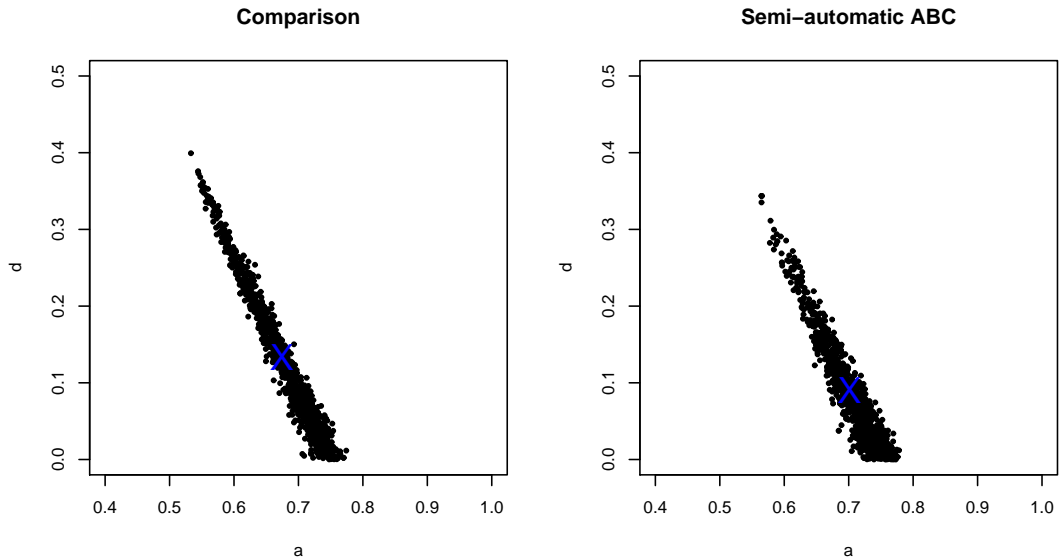}
  \caption{\footnotesize{ABC output for the Tuberculosis application.  Every 1000th state is plotted.  The blue crosses show the ABC means.}}
  \label{fig:TB}
\end{center} \end{figure}

%Summary statistic construction details
As parameters in the pilot output are highly correlated (see Figure \ref{fig:TB}), we fitted a line to the output by linear regression and made a reparameterisation $(u\ v)^T = M (a\ d)^T$ where $M$ is a rotation matrix chosen so that on the fitted line $u$ is constant. The semi-automatic ABC analysis was continued using $(u, v)$ as the parameters of interest.  The explanatory variables $f(\y)$ comprised: the number of clusters of size $i$ for $1 \leq i \leq 5$, the number of clusters of size above 5, average cluster size, $H$, the size of the three largest clusters, and the squares of all these quantities. The semi-automatic ABC analysis used $4 \times 10^6$ simulated datasets in total.

%Results without the truncated prior methodologies are presented, as its only effect was to truncate a small region of the posterior near $d=0$.
Figure \ref{fig:TB} shows the ABC posteriors, indicating that our methodology places less weight on the high $d$ tail of the ABC posterior, reducing the marginal variances from $(0.0029, 0.0088)$ (comparison) to $(0.0017, 0.0048)$ (semi-automatic ABC).

We further investigated this model through a simulation study. We constructed 50 data sets by simulating parameters from the ABC posterior, and simulating data from the model for each of these pairs. We then compared ABC with the summary statistics of \cite{Tanaka:2006} to semi-automatic ABC. Averaged across these data sets, semi-automatic ABC reduced square error loss by around 20\%-25\% for the two parameters. Here we also found that regression correction of \cite{Beaumont:2002} did not change the accuracy of the \cite{Tanaka:2006} method. Using the linear predictors obtained within semi-automatic ABC to estimate parameters (rather than as summary statistics) gave substantially worse estimates, with mean square error for $a$ increasing by a factor of over 3.

%DO WE WANT ALL THE LINEAR PREDICTOR RESUTLS IN?

%To test the robustness of our method we also ran a simulation study on multiple data sets.  We sampled 50 parameter pairs from the comparison ABC posterior shown in Figure \ref{fig:TB} and simulated corresponding datasets.  This ensured data sufficiently similar to the original observations to use exactly the methodology outlined above.  Table \ref{tab:TB} shows that applying semi-automatic ABC reduces mean loss as well as mean variance compared to the comparison analysis.

% latex table generated in R 2.11.0 by xtable 1.5-6 package
% Tue Feb 15 15:01:03 2011
%\begin{table}[htp]
%\begin{center}
%\begin{tabular}{rcc}
%  \hline
% &\multicolumn{2}{c}{Loss} \\
% & $a$ & $d$  \\ 
%  \hline
%  Comparison & 0.0029 & 0.0092  \\ 
%  Comparison + regression & 0.0029 & 0.0094  \\ 
%  Main & {\em 0.0024} & {\em0.0076}  \\ 
%  %Main + regression & 0.0024 & 0.0076  \\ 
%  Linear predictors & 0.0081 & 0.0093  \\ 
%   \hline
%\end{tabular}
%\end{center}
%\label{tab:TB}
%\caption{Mean quadratic losses  of ABC analyses for 50 simulated Tuberculosis datasets}
%\end{table}

\section{Discussion} \label{S:Dis}

We have argued that ABC can be justified by aiming for calibration of the ABC posterior, together with accuracy of estimates of the parameters. We have introduced a new variant of ABC, called noisy ABC, which is calibrated. Standard ABC can be justified as an approximation to noisy ABC, but one which can be more accurate. %The choice between noisy and standard ABC is one of a trade-off between accuracy and calibration. 
Theoretical results suggest that when analysing a single data set, using a relatively small number of summary statistics, that standard ABC should be preferred. 

However we show that when attempting to combine ABC analyses of multiple data sets, noisy ABC is to be preferred. Empirical evidence for this comes from our analysis of a stochastic kinetic network, where using standard ABC within a sequential ABC algorithm, leads to biased parameter estimates. This result could be important for ABC analyses of population genetic data, if inferences are combined across multiple genomic regions. We believe the ideal approach would be to implement a Rao-Blackwellised version of noisy ABC, where we attempt to average out the noise that is added to the summary statistics. If implemented, this would lead to a method which is more accurate than noisy ABC for estimating any function of the parameters. However, at present we are unsure how to implement such a Rao-Blackwellisation scheme in a computationally efficient manner.

The main focus of this paper was a semi-automatic approach to implementing ABC. The main idea is to use simulation to construct appropriate summary statistics, with these summary statistics being estimates of the posterior mean of the parameters. This approach is based on theoretical results which show that choosing summary statistics as the posterior means produces ABC estimators that are optimal in terms of minimising quadratic loss. We have evaluated our method on a number of different models, comparing both to ABC as it has been implemented in the literature, to indirect inference, and the synthetic likelihood approach of \cite{Wood:2010}. 

The most interesting comparison was with between semi-automatic ABC and ABC using the regression correction of \cite{Beaumont:2002}. In a number of examples, these two approaches gave similarly accurate estimates. However, the semi-automatic ABC seemed to be more robust, with the regression correction actually producing less accurate estimates for the Ricker model, and not improving the accuracy for the Tuberculosis model. 

There are alternatives to using linear regression to construct the summary statistics. One approach motivated by our theory, is to use approximate estimates for each parameter. Such an approach has been used in \cite{Wilson/Fearnhead:2009}, where estimates of recombination rate under the incorrect demographic model were used as summary statistics in ABC. It is also the idea behind the approach of \cite{Drovandi/Pettitt/Faddy:2011}, where the authors summarise the data through parameter estimates under an approximating model. A further alternative is to use sliced-inverse regression \cite[]{Li:1991} rather than linear regression to produce the summary statistics. The advantage of sliced-inverse regression is that it is able, where appropriate, to generate multiple linear combinations of the explanatory variables, such that the posterior mean is approximately a function of these linear combinations. Each linear combination could then be used as a summary statistic within ABC.

%Note that our approach can be trivially extended to situations where there are some functions of the parameters that are of primary importance. Assume we have interest in $\phi=\phi(\theta)$, then the optimal choice of summary statistics are the posterior means of the components of $\phi$. If the dimension of $\phi$ is less than that of $\theta$ we could use fewer summary statistics than parameters. This is particularly appealing if there are many nuisance parameters, when implementing ABC using the same number of summary statistics as parameters may not be computationally feasible.  %For the $g$ and $k$ example, we tried this approach by assuming we were interested in only 3 of the 4 parameters, and including one summary statistic for each of these 3 parameters. Our results showed that the overall quadratic loss was almost identical (for those 3 parameters) as when using 4 summary statistics. %An alternative approach for composite likelihood method

Finally we note that there has been much recent research into the computational algorithms underpinning ABC. As well as the rejection sampling, importance sampling and MCMC algorithms we have mentioned, there has been work on using sequential Monte Carlo methods \cite[]{Beaumont:2009,Sisson:2007,Peters:2010}, and adaptive methods \cite[]{Bortot:2007,DelMoral:2009} amongst others. The ideas in this paper are focussing a separate issue within ABC, and we think the semi-automatic approach for implementing ABC that we describe can be utilised within whatever computational method is preferred.

{\bf Acknowledgements:} We thank Chris Sherlock for many helpful discussions, and the reviewers for their detailed comments.
\section*{Appendix}

\subsection*{Appendix A: Variance of IS-ABC and MCMC-ABC}

Calculating the variance of estimates of posterior means using importance sampling are complicated by the dependence between the importance sampling weights and values of $h(\theta)$ \cite[see][]{Liu:1996}. A common approach to quantifying the accuracy of importance sampling is to use an effective sample size. If the weights are normalised to have mean 1, then the effective sample size, $N_{\mbox{eff}}$ is $N$ divided by the mean of the square of the weights. \cite{Liu:1996} argue that for most functions $h(\theta)$, the variance of the estimator in (\ref{eq:1}) will be approximately (\ref{eq:2}) but with $N_{\mbox{acc}}$ replaced by $N_{\mbox{eff}}$.

For a given proposal distribution $g(\theta)$, the effective sample size is
\begin{equation} \label{eq:3b}
N_{\mbox{eff}}=N \frac{\int p(\theta|\Sobs)\pi(\theta) \mbox{d}\theta}{\EABC\left( \pi(\theta)/g(\theta) \right)}. 
\end{equation}
It can be shown that the optimal proposal distribution, in terms of maximising the $N_{\mbox{eff}}/N$  is
\[
g_{\mbox{opt}}(\theta|\yobs) \propto \pi(\theta) p(\theta|\Sobs)^{1/2}.
\]
In which case 
\[
N_{\mbox{eff}}^* = N_{\mbox{acc}} \left(1+\frac{\mbox{Var}_\pi(p(\theta|\Sobs))}{\mbox{E}_\pi(p(\theta|\Sobs)^{1/2})^2} \right),
\]
where the variance and expectation on the right-hand side are with respect to $\pi(\theta)$. It is immediate that $N_{\mbox{eff}}^* \geq N_{\mbox{acc}}$, with equality only if $p(\theta|\Sobs)$ does not depend on $\theta$. The potential gains of importance sampling occur when $p(\theta|\Sobs)$ varies greatly.

Analysis of the Monte Carlo error within the MCMC ABC algorithm is harder. However, consider fixing a proposal kernel $g(\cdot|\cdot)$, which will fix the type of transitions attempted. The Monte Carlo error then will be primarily governed by the
average acceptance probability. For simplicity assume that $K(\cdot)$ is uniform kernel, and that either $g(\cdot|\cdot)$ is chosen to have the prior $\pi(\theta)$ as its stationary distribution, or that the term $ \pi(\theta)g(\theta_{i-1}|\theta)/\{\pi(\theta_{i-1}g(\theta|\theta_{i-1})\} \approx 1$ and can be ignored. The average acceptance probability at stationarity is
\begin{eqnarray*}
 \int\int \piABC(\theta|\Sobs) g(\theta'|\theta) p(\theta'|\Sobs) \mbox{d}\theta\mbox{d}\theta'
&=& \left(\int \pi(\theta)p(\theta|\Sobs)\mbox{d}\theta\right) \times \\
& &  \int\int \piABC(\theta|\Sobs) g(\theta'|\theta) \frac{\piABC(\theta'|\Sobs)}{\pi(\theta')} \mbox{d}\theta\mbox{d}\theta'.
\end{eqnarray*}
The integral comes from averaging over the current and proposed values for the MCMC algorithm, with the average acceptance probability for a given proposed value $\theta'$ being $p(\theta'|\Sobs)$. The right-hand side comes from using (\ref{eq:1a}). The first term on the right-hand side is the average acceptance probability of the rejection algorithm. The second term is 1 if we use an independence sampler $g(\theta'|\theta)$, and will be $>>1$ if the ABC posterior is peaked relative to the prior, and if the transition kernel proposes localised moves.

\subsection*{Appendix B: Proof of Lemma \ref{l:1}}
Write
\begin{eqnarray*}
\int p(\theta|\Sobs)\pi(\theta)\mbox{d}\theta&=& \int \int K(\s-\Sobs)/h) \pi(\s|\theta)\pi(\theta)\mbox{d}\theta\mbox{d} \s \\
& = &  \int h^d K(\x) \pi(\Sobs+h\x) \mbox{d}\x.
\end{eqnarray*}
The first equality comes from the definition of $p(\theta|\Sobs)$, the second by integrating out $\theta$ and making a change of variable $\x=(\s-\Sobs)/h$. So
\[
\left| h^{-d} \int p(\theta|\Sobs)\pi(\theta)\mbox{d}\theta - {\pi(\Sobs)}\right| \leq
\int K(\x) |\pi(\Sobs+h\x)-\pi(\Sobs)|\mbox{d}\x
\]
This bound is shown to be $o(1)$ under either condition.

Consider first condition (i).
Define $c$ to be the maximum value of $|\x|$ such that $K(|\x|)>0$.
For any $\epsilon>0$, by continuity of $\pi(\s)$ at $\Sobs$ we have that there exists a $\delta>0$ such that $|\x|<\delta$ implies $|\pi(\Sobs+\x)-\pi(\Sobs)|<\epsilon$. Define $h_\epsilon=\delta/c$.
Then for $h<h_\epsilon$ we have
\[
 \int K(\x) |\pi(\Sobs+h\x)-\pi(\Sobs)|\mbox{d}\x \leq \epsilon.
\]
This inequality follows as $h|\x|<\delta$ for all $\x$ where $K(\x)>0$. 

Now consider condition (ii).  By differentiable continuity, Taylor's theorem gives
\[
 \pi(\Sobs + h \x) = \pi(\Sobs) + \sum_i h x_i r_i(\x).
\]
The remainder factor $r_i(\x)$ is $| \partial \pi(\mathbf{z})/\partial s_i|$ for some $\mathbf{z}(\x)$,
so by assumption, $|r_i(\x)| \leq R$, a finite bound.  Thus
\begin{eqnarray*}
 \int K(\x) |\pi(\Sobs+h\x)-\pi(\Sobs)|\mbox{d}\x &\leq&
 h R \sum_i \int |x_i| K(\x) \mbox{d}\x.
\end{eqnarray*}

\subsection*{Appendix C: Proof of Theorem \ref{thm:2}}

Rearrangement of the loss function gives
\[
\mbox{E}(L(\theta,\hat{\theta};A)|\yobs) - \mbox{trace}(A\Sigma)
= \mbox{E}((\tilde{\theta}-\hat{\theta})^T A (\tilde{\theta}-\hat{\theta}) | \yobs),
\]
where $\tilde{\theta} = \mbox{E}(\theta|\yobs)$, the mean under the true posterior,
which equals $S(\yobs)$ for the summary statistics under discussion.
Define $\mathbf{\delta}(\Sobs) = \hat{\theta}(\Sobs) - \Sobs$ and make the change of variables $\x=(\tilde{\theta}-\Sobs)/h$.  Now
\begin{eqnarray*}
&&\mbox{E}(L(\theta,\hat{\theta};A)|\yobs) - \mbox{trace}(A\Sigma) - h^2\int \x^T A \x K(\x)\mbox{d}\x \\
&&\quad = -2h\int \x^T A \mathbf{\delta}(\tilde{\theta} + h\x) K(\x)\mbox{d}\x
+\int \mathbf{\delta}(\tilde{\theta} + h\x)^T A \mathbf{\delta}(\tilde{\theta} + h\x) K(\x)\mbox{d}\x.
\end{eqnarray*}
It is required that the modulus of the right hand side is $o(h^2)$.
Let $R$ be the support of $K$. Since this is finite, it suffices to show that
\[
 \max_{\x \in R} \mathbf{\delta}(\tilde{\theta} + h\x) = o(h).
\]
To find an expression for $\mathbf{\delta}$, observe that (\ref{eq:1a}) holds for noisy ABC.
Taking its expectation and making the change of variable $\s = S(\y)$ gives
\[
\hat{\theta}(\mathbf{t}) = \frac{\int \theta \pi(\theta) \pi(\s|\theta) K(\{\s-\mathbf{t}\}/h)
\mbox{d}\theta \mbox{d}\s}{\int \pi(\s) K(\{\s-\mathbf{t}\}/h) \mbox{d}\s}.
\]
Note
\[
\int \theta \pi(\theta) \pi(\s|\theta) \mbox{d} \theta = \pi(\s) \mbox{E}(\theta|\s) = \s \pi(s),
\]
where the second equality is due to our choice of $S$.
Thus
\[
\mathbf{\delta}(\mathbf{t}) = \frac{\int (\s-\mathbf{t}) \pi(\s) K(\{\s-\mathbf{t}\}/h) \mbox{d}\s}
{\int \pi(\s) K(\{\s-\mathbf{t}\}/h) \mbox{d}\mathbf{s}}.
\]
Make the change of variables $\mathbf{y}=(\s-\mathbf{t})/h$
and consider the case $\mathbf{t}=\tilde{\theta} + h\x$, then
\[
\mathbf{\delta}(\tilde{\theta} + h\x) = \frac{h \int \mathbf{y} \pi(\tilde{\theta}+h[\x+\mathbf{y}]) K(\mathbf{y}) \mbox{d}\mathbf{y}} {\int \pi(\tilde{\theta}+h[\x+\mathbf{y}]) K(\mathbf{y}) \mbox{d}\mathbf{y}}.
\]
By the argument in Appendix B, continuity of $\pi(\s)$ at $\s=\tilde{\theta}$
gives denominator $\pi(\tilde{\theta}) + o(1)$.
Consider the $i$th component of the integral in the numerator
\begin{eqnarray*}
&& \left| \int y_i \pi(\tilde{\theta}+h[\x+\mathbf{y}]) K(\mathbf{y}) \mbox{d}\mathbf{y} -
\int y_i \pi(\tilde{\theta}) K(\mathbf{y}) \mbox{d}\mathbf{y} \right| \\
&\leq& \max_{y \in R} |y_i| \left| \int (\pi(\tilde{\theta}+h[\x+\mathbf{y}]) - \pi(\tilde{\theta})) K(\mathbf{y}) \mbox{d}\mathbf{y} \right|
\end{eqnarray*}
This integral is $o(h)$ by the continuity argument just mentioned.
Noting that $\int y_i K(\mathbf{y}) \mbox{d}\mathbf{y} = 0$ by assumption, we have
\[
\int \mathbf{y} \pi(\tilde{\theta}+h[\x+\mathbf{y}]) K(\mathbf{y}) \mbox{d}\mathbf{y} = o(1).
\]
Combining these results gives the required bound for $\delta$.
\bibliography{dennis}

\end{document}